\documentclass[english,draftclsnofoot,onecolumn,11pt]{IEEEtran}
\usepackage[T1]{fontenc}
\usepackage{verbatim}
\usepackage{amsthm}
\usepackage{amsmath}
\usepackage{graphicx}
\usepackage{amssymb}

\makeatletter
\theoremstyle{plain}
\newtheorem{thm}{Theorem}
\theoremstyle{definition}
\newtheorem{example}[thm]{Example}
\theoremstyle{plain}
\newtheorem{lem}[thm]{Lemma}
\theoremstyle{remark}
\newtheorem{rem}[thm]{Remark}
\theoremstyle{definition}
\newtheorem{defn}[thm]{Definition}
\theoremstyle{plain}
\newtheorem{cor}[thm]{Corollary}
\theoremstyle{plain}
\newtheorem{conjecture}[thm]{Conjecture}

\usepackage{bbm}
\usepackage{multicol}
\usepackage{amsfonts}
\usepackage{euscript}

\DeclareMathOperator*{\sgn}{sgn}
\DeclareMathOperator*{\argmin}{arg\,min}
\DeclareMathOperator*{\argmax}{arg\,max}

\IEEEoverridecommandlockouts

\newcommand{\op}[1]{\mathsf{#1}}
\newcommand{\set}[1]{\mathcal{#1}}
\newcommand{\vect}[1]{\boldsymbol{#1}}

\makeatother

\usepackage{babel}

\begin{document}

\title{Convergence of Weighted Min-Sum Decoding Via Dynamic Programming
on Trees}

\author{Yung-Yih Jian and Henry D. Pfister\thanks{This work was supported in part by the National Science Foundation under Grant No. 0747470 and the Texas Norman Hackerman Advanced Research Program under  Grant No. 000512-0168-2007. Any opinions, findings, conclusions, or recommendations expressed in this material are those of the authors and do not necessarily reflect the views of the sponsors.}\\
Department of Electrical and Computer Engineering, Texas A\&M University\\
Email: \{yungyih.jian,hpfister\}@tamu.edu}
\maketitle
\begin{abstract}
Applying the max-product (and belief-propagation) algorithms to loopy
graphs is now quite popular for best assignment problems. This is
largely due to their low computational complexity and impressive performance
in practice. Still, there is no general understanding of the conditions
required for convergence and/or the optimality of converged solutions.
This paper presents an analysis of both attenuated max-product (AMP)
decoding and weighted min-sum (WMS) decoding for LDPC codes which
guarantees convergence to a fixed point when a weight parameter, $\beta$,
is sufficiently small. It also shows that, if the fixed point satisfies
some consistency conditions, then it must be both the linear-programming
(LP) and maximum-likelihood (ML) solution. 

For $(d_{v},d_{c})$-regular LDPC codes, the weight must satisfy $\beta(d_{v}-1)\leq1$
whereas the results proposed by Koetter and Frey require instead that
$\beta(d_{v}-1)(d_{c}-1)<1$. A counterexample which shows a fixed
point might not be the ML solution if $\beta(d_{v}-1)>1$ is also
given. Finally, connections are explored with recent work by Arora
et al. on the threshold of LP decoding.\end{abstract}

\vspace{-1mm}

\section{Introduction}

The introduction of turbo codes in 1993 started a revolution in coding
and inference that continued with the rediscovery of low-density parity-check
(LDPC) codes and culminated in optimized LDPC codes that essentially
achieve the capacity of practical channels \cite{Berrou-icc93,Gallager-1963,Mackay-it99,Richardson-it01*2}.
During this time, Wiberg \emph{et al.} advanced the analysis of iterative
decoding by proving a number of results for the min-sum (a.k.a. max-product)
decoding algorithm \cite{Wiberg-96,Wiberg-ett95,Frey-amfm00}. Richardson
and Urbanke also introduced the technique of density evolution (DE)
to compute noise thresholds of message-passing decoding algorithms
for turbo and LDPC codes \cite{Richardson-it01}. 

For a particular noise realization, the optimality of iterative decoding
solutions has also been considered by a number of authors. Weiss and
Freeman have shown that the max-product (MP) assignment is locally
optimal w.r.t. all single-loop and tree perturbations \cite{Weiss-it01}.
Unfortunately, this result is typically uninformative for LDPC codes
with variables degrees larger than 2. Frey and Koetter have also shown
that, with proper weights and adjustments, the attenuated max-product
(AMP) decoding for LDPC codes returns the maximum-likelihood (ML)
codeword if it converges to a codeword \cite{Frey-amfm00}. For general
graphs, Wainwright \emph{et al. }proposed the tree-reweighted max-product
(TRMP) message-passing algorithm for computing the MAP assignment
on the strictly positive Markov random field \cite{Wainwright-it05}.
They have shown that, under some optimality conditions, the converged
solution gives the MAP configuration for the graph. Their algorithm,
though strictly different, has some similarity to the AMP algorithm
in \cite{Frey-amfm00}. 

The linear programming (LP) decoding for LDPC codes, proposed by Feldman
\emph{et al.}, solves a relaxed version of the ML decoding problem
\cite{Feldman-it05}. Since its introduction, a number of authors
have looked for connections to the MP iterative decoding algorithm
\cite{Koetter-istc03}. One interesting open question is, {}``What
is the noise threshold of LP decoding?''. The first threshold bound
for LP decoding was proposed in \cite{Feldman-it07}. Using expander
graph arguments, they showed that LP decoding of a rate-$\frac{1}{2}$
regular LDPC code can correct all error patterns with weight less
than $0.000175$ of the block length. Since this is a worst-case analysis,
the large gap to the empirical observations is not too surprising.
Daskalakis \emph{et al. }\cite{Daskalakis-it08} were able to improve
the threshold to $0.002$ using probabilistic arguments based on a
construction of a LP dual feasible solution. In \cite{Koetter-ita06},
Koetter and Vontobel applied girth-based arguments to the dual LP
problem. For a $(3,6)$-regular LDPC code, they proved that LP decoding
can tolerate a crossover probability of $p=0.01$ on the binary symmetric
channel (BSC) and a noise level of $\sigma=0.5574$ on the binary-input
additive white Gaussian noise channel (BIAWGNC). 

Arora \emph{et al.} showed recently that, for a $(3,6)$-regular LDPC
code, LP decoding can tolerate a crossover probability $p=0.05$ on
the BSC \cite{Arora-stoc09}. Instead of using a dual LP solution,
they investigated the primal solution of the LP problem and proposed
a local optimality condition for codewords. They proved that the local
optimality implies both global optimality and LP optimality. So the
probability that LP decoding succeeds is lower bounded by the probability
that the correct codeword satisfies a set of local optimality conditions.
Since their local optimality conditions are amenable to analysis on
tree-like neighborhoods, they perform a DE analysis to obtain BSC
noise thresholds for LP decoding. Using DE for memoryless binary-input
output symmetric (MBIOS) channels, Halabi \emph{et al. }showed that
LP decoding can achieve a noise threshold $\sigma=0.735$ on the BIAWGNC
\cite{Halabi-arxiv10}.

The results in this paper can be seen as an extension of the work
by Frey and Koetter that provides new insight into results of \cite{Koetter-ita06,Arora-stoc09}.
We view both AMP and WMS \cite{Chen-comlett02} algorithms as computing
the dynamic-programming (DP) solution to the optimal discounted-reward
problem on a set of overlapping trees. This allows us to show that,
for any received vector, the one-step update of the algorithm is a
contraction on the space of message values when weight parameter is
sufficiently small. From this, we deduce that the messages converge
to a unique fixed point. We first show that, for $(d_{v},d_{c})$-regular
LDPC codes, if the resulting fixed point satisfies some consistency
conditions, then it must also be the LP optimum solution and, hence,
the ML solution. Then, the WMS algorithm on $(d_{v},d_{c})$-regular
LDPC codes with messages diverging to $\pm\infty$ is considered.
We show that, for the weight $\beta=\frac{1}{d_{v}-1}$, if the WMS
messages diverge to $\pm\infty$ and satisfies the consistency conditions,
the corresponding hard decisions also return the ML solution.

The rest of the paper is organized as follows. Section \ref{sec:Dynamic-Programming}
provides the background on factor graphs as well as the update rules
of the AMP algorithm and the WMS algorithm. In Section \ref{sub:Convergence-and-Optimality},
we first investigate the convergence property of both algorithms,
and introduce the consistency conditions for both algorithms. Then,
the optimality of the hard decisions corresponding to the consistent
fixed point is discussed. In Section \ref{sec: BetaEq1/(dv-1)}, the
optimality of the codeword returned by the WMS algorithm when the
messages are not converged is analyzed. A conjecture about the connections
between noise thresholds of the WMS decoding and noise thresholds
of the LP decoding is proposed in the same section. Numerical results
are described and discussed in Section \ref{sec:Numerical-Results}.
Finally, conclusions and extensions are given in Section \ref{sec:Conclusions-and-Future}.

\section{Background\label{sec:Dynamic-Programming}}

\subsection{Factor Graphs}

An LDPC code can be defined by a bipartite graph $\set{G}=\left(\set{V},\set{E}\right)$,
where $\set{E}$ is the set of edges, and $\set{V}=\set{V}_{L}\cup\set{V}_{R}$
consists of variable nodes (or bit nodes) $\set{V}_{L}$ and check
nodes (or constraint nodes) $\set{V}_{R}$. In this paper, $(d_{v},d_{c})$-regular
LDPC codes are considered. That is, each variable node in $\set{V}_{L}$
has $d_{v}$ edges attached to it, and each check node in $\set{V}_{R}$
has $d_{c}$ edges attached to it. For a set $\set{S}$, let $|\set{S}|$
denote the cardinality of $\set{S}$. The number of variable nodes
denoted by $n$ is $|\set{V}_{L}|$. Any binary vector $\boldsymbol{x}\in\{0,1\}^{n}$
is a codeword, or a valid assignment, if and only if it satisfies
all check nodes in $\set{V}_{R}$. We use $\set{C}$ to denote the
collection of all codewords. Let $\mathcal{T}_{i}^{L}$ be a computation
tree of $\set{G}$ which has depth $L$ and is rooted at node $i$.
The set of vertices in the $\ell$th level of $\set{T}_{i}^{L}$,
where $\ell\leq L$, is denoted by $N(i,\ell)$. We also consider
computation trees of $\set{G}$ rooted at a directed edge $i\rightarrow j$
or $i\leftarrow j$ with $(i,j)\in\set{E}$. For a graph $\set{G}$,
the size of the smallest cycle in $\set{G}$ is denoted by girth($\set{G}$).
For a node $i\in\set{V}$, we use $N(i)$ to denote the set of neighbors
of $i$.

\subsection{Discounted Dynamic Programming on a Tree\label{sub:Discounted-Dynamic-Programming}}

Suppose that the computation tree $\mathcal{T}_{i}^{2L}$ has depth
$2L<\frac{1}{2}\mbox{girth}(\set{G})$. Then, each node in $\mathcal{T}_{i}^{2L}$
is associated with a different node in $\set{G}$. Let $\mathcal{I}\subset\set{V}_{L}$
be the subset of variable nodes in $\mathcal{T}_{i}^{2L}$. A binary
vector $\vect{w}\in\{0,1\}^{n}$ is a valid assignment on $\mathcal{T}_{i}^{2L}$
if $\vect{w}$ satisfies all check nodes in $\set{T}_{i}^{2L}$. Let
$\mathcal{C}_{\mathcal{T}_{i}^{2L}}$ be the set of all valid assignments
on $\mathcal{T}_{i}^{2L}$, and let $\mathcal{C}_{\mathcal{T}_{i}^{2L}}(x)\triangleq\{\vect{w}\in\mathcal{C}_{\mathcal{T}_{i}^{2L}}:w_{i}=x,w_{m}=0,\forall m\notin\set{T}_{i}^{2L}\}$
be a subset of $\mathcal{C}_{\mathcal{T}_{i}^{2L}}$, where the assignment
of the node $m\in\set{V}_{L}\setminus\set{I}$ is $0$ and the assignment
of the root node is $x$. In the remainder of this paper, we often
simplify $\set{C}_{\set{T}_{i}^{2L}}$ to $\set{C}_{\set{T}}$ when
$i$ and $L$ are evident from the context. Similarly, we also simplify
$\set{C}_{\set{T}_{i}^{2L}}(x)$ to $\set{C}_{\set{T}}(x)$. 

Let $\gamma_{i}(x_{i})\triangleq\log(p_{Y|X}(y_{i}|x_{i}))$ be the
log-likelihood of receiving $y_{i}\in\mathbb{R}$ given that $x_{i}\in\{0,1\}$
is transmitted at the $i$th bit. We consider the problem of finding
the best assignment $\boldsymbol{w}^{*}\in\{\mathcal{C_{T}}(0)\cup\mathcal{C_{T}}(1)\}$
to a tree $\mathcal{T}_{i}^{2L}$ defined by \begin{align}
\boldsymbol{w}^{*} & \triangleq\argmax_{\boldsymbol{w}\in\{\mathcal{C_{T}}(0)\,\cup\,\mathcal{C_{T}}(1)\}}\sum_{m\in\mathcal{I}}\beta_{m}\gamma_{m}(w_{m}),\label{eq: amp org def}\end{align}
where $\beta_{m}=\beta^{\ell}$ if $m\in N(i,2\ell)$ and $0$ otherwise.
Since $\set{C_{T}}(0)$ and $\set{C_{T}}(1)$ are disjoint, (\ref{eq: amp org def})
can be separated into two subproblems. For $x\in\{0,1\}$, we first
define a vector\begin{equation}
\boldsymbol{w}^{*}(x)\triangleq\argmax_{\boldsymbol{w}\in\mathcal{C_{T}}(x)}\sum_{m\in\mathcal{I}}\beta_{m}\gamma_{m}(w_{m}),\label{eq: w*x}\end{equation}
and a function \begin{align}
\mu_{i}(x) & \triangleq\max_{\boldsymbol{w}\in\mathcal{C_{T}}(x)}\sum_{m\in\mathcal{I}}\beta_{m}\gamma_{m}(w_{m}),\label{eq: reward func}\end{align}
where $\mu_{i}(x)$ is the optimal reward for assigning $x$ to the
root node of $\mathcal{T}_{i}^{2L}$, and $\vect{w}^{*}(x)$ is the
corresponding best assignment. Then, the solution of (\ref{eq: amp org def})
is obtained by choosing $\vect{w}^{*}(x^{*})$ where $x^{*}=\argmax_{x\in\{0,1\}}\mu_{i}(x)$.
Note that $\mu_{i}(\cdot)$ is only a function of the assignment to
the root node of $\mathcal{T}_{i}^{2L}$. Therefore, finding the best
assignment of the tree $\set{T}_{i}^{2L}$ is equivalent to finding
the best assignment of the root node of $\set{T}_{i}^{2L}$.

The RHS in (\ref{eq: reward func}) can be rewritten as\begin{align}
\mu_{i}(x) & =\gamma_{i}(x)+\max_{\boldsymbol{w}\in\mathcal{C}_{\mathcal{T}}(x)}\sum_{\ell=1}^{L}\sum_{m\in N(i,2\ell)}\beta^{\ell}\gamma_{m}(w_{m}).\label{eq: amp reward func}\end{align}
This suggests that we can compute $\mu_{i}(x)$ recursively by using
DP. In the $(\ell+1)$th iteration, we compute the optimal discounted
total reward $\mu_{i\rightarrow j}^{(\ell+1)}(x)$ of assigning $x$
to the directed edge $i\rightarrow j$ by \begin{align}
\mu_{i\rightarrow j}^{(\ell+1)}(x) & =\gamma_{i}(x)+\beta\sum_{k\in N(i)\backslash j}\mu_{i\leftarrow k}^{(\ell)}(x)\nonumber \\
 & =\gamma_{i}(x)+\beta\sum_{k\in N(i)\backslash j}\max_{\boldsymbol{w}\in\mathcal{S}_{k,i}(x)}\sum_{m\in N(k)\backslash i}\mu_{m\rightarrow k}^{(\ell)}(w_{m}),\label{eq: AMPv2c}\end{align}
where\begin{equation}
\mathcal{S}_{k,i}(x)\triangleq\Big\{\boldsymbol{w}\in\{0,1\}^{n}:\, w_{i}=x,\,\sum_{m\in N(k)}w_{m}=0\mod2\Big\}\label{eq: LocalPolytopeVertices}\end{equation}
is the set of all valid assignments for variables in constraint $k$
when $x$ is assigned to the directed edge $i\leftarrow k$. This
follows from defining $\mu_{i\leftarrow k}^{(\ell)}(x)$ to be the
optimal discounted total reward for assigning $x$ to the directed
edge $i\leftarrow k$ according to the rule

\begin{equation}
\mu_{i\leftarrow k}^{(\ell)}(x)=\max_{\boldsymbol{w}\in\mathcal{S}_{k,i}(x)}\sum_{m\in N(k)\backslash i}\mu_{m\rightarrow k}^{(\ell)}(w_{m}).\label{eq: AMPc2v}\end{equation}
Finally, the reward function (\ref{eq: amp reward func}) can be computed
by \[
\mu_{i}(x)=\gamma_{i}(x)+\beta\sum_{j\in N(i)}\mu_{i\leftarrow j}^{(L)}(x).\]
To initialize the process, we choose $\mu_{i\rightarrow j}^{(0)}(x)=\gamma_{i}(x)$
for all edges $i\rightarrow j$ and all $x\in\{0,1\}$. The update
rule in (\ref{eq: AMPv2c}) is the same as the AMP algorithm proposed
in \cite{Frey-amfm00}, where the optimal discounted total rewards
$\mu_{i\rightarrow j}^{(\ell)}(x)$ and $\mu_{i\leftarrow j}^{(\ell)}(x)$
are the messages passed on the directed edges $i\rightarrow j$ and
$i\leftarrow j$, respectively.

By using the update rule (\ref{eq: AMPv2c}), one can compute $\mu_{i}(x)$
for all $i\in\set{V}_{L}$ in parallel. Suppose that the total number
of iterations $L$ is less than $\frac{1}{4}\mbox{girth(}\set{G})$.
The vector $\vect{x}^{*}$ with $x_{i}^{*}=\argmax_{x\in\{0,1\}}\mu_{i}(x)$
is the best assignments of the root of the trees $\{\set{T}_{i}^{2L}:i\in\set{V}_{L}\}$.
Since $\set{T}_{i}^{2L}$ for all $i\in\set{V}_{L}$ are overlapped,
any variable node $i\in\set{V}_{L}$ appears in more than one tree.
In both \cite{Frey-amfm00} and \cite{Arora-stoc09}, it has been
shown that if the best assignment of each $i\in\set{V}_{L}$ is consistent
across all trees, then $\vect{x}^{*}$ is the ML solution. To check
the optimality of $\vect{x}^{*}$, one has to first find the best
assignment $\vect{w}^{*}$ of each computation tree, and then test
whether the assignment $\vect{w}^{*}$ of each tree is consistent
with $\vect{x}^{*}$ or not. In this paper, we discuss how the weight
factor $\beta$ affects the decoder. We also propose other consistency
conditions, which are easier to check, for regular LDPC codes. Finally,
the analysis is extended to $L>\frac{1}{4}\mbox{girth}(\set{G})$.

\subsection{Attenuated Max-product Decoding Algorithm}

In Section \ref{sub:Discounted-Dynamic-Programming}, the original
AMP algorithm was introduced. In this section, we introduce a modified
version of the AMP algorithm, which is mathematically equivalent to
the original one for any finite number of iterations.

Let $\gamma_{i}\triangleq\gamma_{i}(0)-\gamma_{i}(1)$ be the channel
log-likelihood ratio (LLR) for the $i$th bit. It can be shown that
$\boldsymbol{w}^{*}(x)$ defined in (\ref{eq: w*x}) also maximizes
the following objective function \begin{align}
\mu_{i}(x) & =\sum_{\vect{w}\in\mathcal{C_{T}}(x),\, m\in\mathcal{I}}\beta_{m}(1-w_{m})\gamma_{m}.\label{eq: mod amp}\end{align}
To show the equivalence between the objective function of (\ref{eq: w*x})
and (\ref{eq: mod amp}), we subtract a constant $\sum_{m\in\mathcal{I}}\beta_{m}\gamma_{m}(1)$
from the objective function of (\ref{eq: w*x}). Then,\begin{align*}
\argmax_{\boldsymbol{w}\in\mathcal{C_{T}}(x)}\sum_{m\in\mathcal{I}}\beta_{m}\gamma_{m}(w_{m}) & =\argmax_{\boldsymbol{w}\in\mathcal{C_{T}}(x)}\sum_{m\in\mathcal{I}}\beta_{m}\gamma_{m}(w_{m})-\sum_{m\in\mathcal{I}}\beta_{m}\gamma_{m}(1)\\
 & =\argmax_{\boldsymbol{w}\in\mathcal{C_{T}}(x)}\sum_{m\in\mathcal{I}}\beta_{m}\left(\gamma_{m}(w_{m})-\gamma_{m}(1)\right)\\
 & =\argmax_{\boldsymbol{w}\in\mathcal{C_{T}}(x)}\sum_{m\in\mathcal{I}}\beta_{m}\left(1-w_{m}\right)\gamma_{m}.\end{align*}
Therefore, the modified AMP update rule becomes \begin{align}
\mu_{i\rightarrow j}^{(\ell+1)}(x) & =(1-x)\gamma_{i}+\beta\sum_{k\in N(i)\backslash j}\mu_{i\leftarrow k}^{(\ell)}(x)\nonumber \\
 & =(1-x)\gamma_{i}+\beta\sum_{k\in N(i)\backslash j}\max_{\boldsymbol{w}\in\mathcal{S}_{k,i}(x)}\sum_{m\in N(k)\backslash i}\mu_{m\rightarrow k}^{(\ell)}(w_{m}),\label{eq: amp mu_ij update}\end{align}
where the message $\mu_{i\rightarrow j}^{(\ell+1)}(x)$ now represents
the weighted correlation between the LLRs and the best valid assignment
with $x$ assigned to the directed edge $i\rightarrow j$. The algorithm
starts by setting $\mu_{i\rightarrow j}^{(0)}(x)=(1-x)\gamma_{i}$
for all $(i,j)\in\set{E}$ and all $x\in\{0,1\}$.

Similar to the analysis in Section \ref{sub:Discounted-Dynamic-Programming},
$\mu_{i\rightarrow j}(x)$ can be considered as a DP value function,
that assigns a real number to each bit-to-check directed edge $i\rightarrow j$
and each possible assignment $x\in\{0,1\}$. Based on the standard
approach to DP, the update process can be seen as applying an operator
$\mathsf{A}:\,\mathbb{R}^{2|\set{E}|}\rightarrow\mathbb{R}^{2|\set{E}|}$
to messages. Let $\boldsymbol{\mu}\in\mathbb{R}^{2|\set{E}|}$ with
$\boldsymbol{\mu}\triangleq\{\mu_{i\rightarrow j}(x):(i,j)\in\set{E},x\in\{0,1\}\}$
be an AMP message vector. From (\ref{eq: amp mu_ij update}), the
operator $\mathsf{A}$ is defined by $\boldsymbol{\nu}=\mathsf{A}[\boldsymbol{\mu}]$
with\begin{align}
\nu_{i\rightarrow j}(x) & =(1-x)\gamma_{i}+\beta\sum_{k\in N(i)\backslash j}\mu_{i\leftarrow k}(x).\nonumber \\
 & =(1-x)\gamma_{i}+\beta\sum_{k\in N(i)\backslash j}\max_{\boldsymbol{w}\in\mathcal{S}_{k,i}(x)}\sum_{m\in N(k)\backslash i}\mu_{m\rightarrow k}(w_{m}).\label{eq: DefOperatorA}\end{align}
The AMP algorithm proceeds iteratively by computing $\boldsymbol{\mu}^{(\ell+1)}=\mathsf{A}[\boldsymbol{\mu}^{(\ell)}]$.

\subsection{Weighted Min-sum Decoding Algorithm}

Instead of passing the vector $(\mu_{i\rightarrow j}^{(\ell)}(0),\,\mu_{i\rightarrow j}^{(\ell)}(1))\in\mathbb{R}^{2}$
as the $i\rightarrow j$ message in the AMP algorithm, the WMS algorithm
passes message $\mu_{i\rightarrow j}^{(\ell)}\triangleq\mu_{i\rightarrow j}^{(\ell)}(0)-\mu_{i\rightarrow j}^{(\ell)}(1)$,
which is simply the difference between the best 0-root correlation
and the best 1-root correlation. Similarly, the $i\leftarrow j$ message
is simplified to $\mu_{i\leftarrow j}^{(\ell)}\triangleq\mu_{i\leftarrow j}^{(\ell)}(0)-\mu_{i\leftarrow j}^{(\ell)}(1)$.
The update rules of the WMS algorithm are therefore given by\begin{align}
\mu_{i\rightarrow j}^{(\ell+1)} & =\gamma_{i}+\beta\sum_{k\in N(i)\setminus j}\mu_{i\leftarrow k}^{(\ell)},\label{eq: WMS-v2c}\\
\mu_{i\leftarrow j}^{(\ell)} & =\left(\prod_{m\in N(j)\setminus i}\sgn\left(\mu_{m\rightarrow j}^{(\ell)}\right)\right)\min_{m'\in N(j)\setminus i}\left|\mu_{m'\rightarrow j}^{(\ell)}\right|.\label{eq: WMS-c2v}\end{align}
 It is easy to verify that the WMS algorithm is equivalent to the
AMP algorithm. 

Similar to the AMP algorithm, for any WMS message vector $\boldsymbol{\mu}\in\mathbb{R}^{\left|\set{E}\right|}$
with $\vect{\mu}\triangleq\{\mu_{i\rightarrow j}:(i,j)\in\set{E}\}$,
the update rule of the WMS algorithm can be seen as an operator $\mathsf{W}:\mathbb{R}^{\left|\set{E}\right|}\rightarrow\mathbb{R}^{\left|\set{E}\right|}$,
which is defined by $\boldsymbol{\nu}=\mathsf{W}[\boldsymbol{\mu}]$
with\begin{align}
\nu_{i\rightarrow j} & =\gamma_{i}+\beta\sum_{k\in N(i)\setminus j}\left(\prod_{m\in N(k)\setminus i}\sgn\left(\mu_{m\rightarrow k}\right)\right)\min_{m'\in N(k)\setminus i}\left|\mu_{m'\rightarrow k}\right|.\label{eq: WMSOperator}\end{align}
The WMS algorithm is initialized by setting $\mu_{i\rightarrow j}^{(0)}=\gamma_{i}$
and proceeds iteratively by computing $\boldsymbol{\mu}^{(\ell+1)}=\op{W}[\boldsymbol{\mu}^{(\ell)}]$.

\subsection{LP Decoding}

Given the received vector $\boldsymbol{y}\in\mathbb{R}^{n}$, the
ML decoder finds a codeword $\boldsymbol{x}^{*}\in\mathcal{C}$ such
that the probability $p(\boldsymbol{y}|\boldsymbol{x}^{*})$ is maximal
among all $\boldsymbol{x}\in\mathcal{C}$. Let $\boldsymbol{\gamma}\in\mathbb{R}^{n}$
be the vector of channel LLRs. Then, ML decoding can be defined as
the following integer programming problem \cite{Feldman-it05}, \begin{equation}
\begin{array}{llc}
\mbox{min} & \sum_{i=1}^{n}\gamma_{i}x_{i}\\
\mbox{subject to} & \boldsymbol{x}\in\mathcal{C}.\end{array}\label{eq: ML LLR form 2}\end{equation}
For a fixed graph $\set{G}$, solving (\ref{eq: ML LLR form 2}) directly
is computationally infeasible for large $n$ because the number of
codewords grows exponentially in $n$. In \cite{Feldman-it05}, a
suboptimal decoder, i.e., LP decoder, was proposed. With the same
objective function as in (\ref{eq: ML LLR form 2}), the LP decoder
searches the optimal solution over a relaxed polytope which is obtained
by intersecting all local codeword polytopes defined by each check
node of the graph $\set{G}$. 

Here, we briefly describe the LP decoder in \cite{Feldman-it05} as
follows. Given a check node $j\in\set{V}_{R}$, let \[
\mathcal{E}_{j}=\left\{ S\subseteq N(j):\,|S|\mbox{ is even}\right\} \]
be the collection of all support sets of local codewords for $j$.
Note that $\emptyset\in\mathcal{E}_{j}$ and represents the all-zeros
codeword. For each $j\in\set{V}_{R}$, and $S\in\mathcal{E}_{j}$,
$\zeta_{j,S}$ is an indicator function of the local codeword being
assigned to $j$. The LP decoder solves the following problem

\[
\begin{array}{lll}
\mbox{min} & {\displaystyle \sum_{i\in\set{V}_{L}}\gamma_{i}x_{i}}\\
\mbox{subject to} & {\displaystyle \sum_{S\in\mathcal{E}_{j}}\zeta_{j,S}=1} & \forall j\in\set{V}_{R}\\
 & {\displaystyle \sum_{\substack{S\in\set{E}_{j}\\
S\ni i}
}\zeta_{j,S}=x_{i}} & \forall(i,j)\in\set{E}\\
 & {\displaystyle \zeta_{j,S}\geq0},\ {\displaystyle x_{i}\geq0} & \forall i\in\set{V}_{L},\ \forall j\in\set{V}_{R},\ \forall S\in\mathcal{E}_{j}.\end{array}\]
If the solution vector $\vect{x}^{*}$ is in $\{0,1\}^{n}$, then
the vector $\boldsymbol{x}^{*}$ is an ML codeword. In the sequel,
this LP problem is called \emph{Problem-P}.

To establish the dual problem of \emph{Problem-P}, a Lagrange multiplier
$\tau_{i,j}$ is associated with each edge $(i,j)\in\set{E}$ of the
graph $\set{G}$. The resulting dual problem is given by \[
\begin{array}{llc}
\mbox{max} & {\displaystyle \sum_{j\in\set{V}_{R}}\tau_{j}}\\
\mbox{subject to} & {\displaystyle \sum_{i\in S}\tau_{i,j}\geq\tau_{j}} & \forall j\in\set{V}_{R},\ \forall S\in\mathcal{E}_{j}\\
 & {\displaystyle \sum_{j\in N(i)}}\tau_{i,j}\leq\gamma_{i} & \forall i\in\set{V}_{L},\end{array}\]
which, as shown in \cite{Koetter-ita06}, is equivalent to \[
\begin{array}{llc}
\mbox{max} & {\displaystyle \sum_{j\in\set{V}_{R}}\min_{S\in\mathcal{E}_{j}}\sum_{i\in S}\tau_{i,j}}\\
\mbox{subject to} & {\displaystyle \sum_{j\in N(i)}}\tau_{i,j}=\gamma_{i} & \forall i\in\set{V}_{L}.\end{array}\]
In the remainder of this paper, this dual problem is called \emph{Problem-D}. 

Consider a $(d_{v},d_{c})$-regular LDPC code, and let \begin{equation}
\set{L}\triangleq\left\{ \boldsymbol{w}\in\{0,1\}^{d_{c}}:\sum_{i=1}^{d_{c}}w_{i}=0\mod2\right\} \label{eq: ListOfLocalCodeword}\end{equation}
be the set of locally valid codewords. For each check node $j\in\set{V}_{R}$,
we define a vector $\boldsymbol{\tau}_{j}=\{\tau_{i,j}:i\in N(j)\}$.
Then the objective function in \emph{Problem-D} can be written as
$\sum_{j\in\set{V}_{R}}\min_{\boldsymbol{w}\in\set{L}}\left\langle \boldsymbol{w},\boldsymbol{\tau}_{j}\right\rangle $.

\subsection{Impossibility of a General ML Certificate for WMS Decoding}

%
{} In this section, two examples are provided for showing that WMS algorithm
with some $\beta>\frac{1}{d_{v}-1}$ is not guaranteed to return an
ML codeword.
\begin{example}
In this example, the ML optimality of the codeword returned by the
WMS decoder with $\beta=0.8$ is checked. We consider a $(3,4)$-regular
LDPC code over the BSC channel with cross-over probability $p=0.1$.
The parity check matrix for the $(3,4)$-regular LDPC code is\begin{align*}
H & =\left(\begin{array}{c}
0,\,0,\,0,\,1,\,1,\,0,\,1,\,0,\,0,\,0,\,0,\,1\\
0,\,0,\,0,\,0,\,0,\,1,\,1,\,1,\,0,\,1,\,0,\,0\\
0,\,1,\,0,\,1,\,0,\,1,\,0,\,0,\,1,\,0,\,0,\,0\\
1,\,1,\,1,\,0,\,0,\,0,\,1,\,0,\,0,\,0,\,0,\,0\\
1,\,0,\,0,\,1,\,0,\,0,\,0,\,0,\,0,\,1,\,1,\,0\\
0,\,0,\,1,\,0,\,1,\,1,\,0,\,0,\,0,\,0,\,1,\,0\\
1,\,0,\,0,\,0,\,1,\,0,\,0,\,1,\,1,\,0,\,0,\,0\\
0,\,1,\,0,\,0,\,0,\,0,\,0,\,1,\,0,\,0,\,1,\,1\\
0,\,0,\,1,\,0,\,0,\,0,\,0,\,0,\,1,\,1,\,0,\,1\end{array}\right).\end{align*}
Since the codeword length is short ($n=12$), we are able to implement
the ML decoder defined in (\ref{eq: ML LLR form 2}). For the WMS
decoder, $200$ iterations are performed in decoding each block. After
testing $10^{5}$ blocks, there are $90905$ codewords returned by
the WMS decoder. Among these codewords returned by the WMS algorithm,
only $90850$ codewords are the ML codeword. Therefore, codewords
returned by the WMS algorithm with $\beta=0.8$ cannot be guaranteed
to be ML optimal. 
\end{example}

For the general case, the following example gives some intuition.
\begin{example}
Consider a $(d_{v},d_{c})$-regular LDPC code with codeword length
$n$, where $d_{c}$ is an odd number and $d_{v}>3.$ Assume the all-zeros
codeword is transmitted. Let the channel output LLR be $\boldsymbol{\gamma}=(-1,\dots,-1).$
Consider the WMS algorithm with $\beta>\frac{2}{d_{v}-1}.$ 

At the beginning, all messages from variable nodes to their neighboring
check nodes are $\mu_{i\rightarrow j}^{(0)}=-1$ for $i=1,\dots,n$
and $j\in N(i).$ Consider the message passed from the $j$th check
nodes to its neighbor variable nodes, $\mu_{i\leftarrow j},\ i\in N(j).$
Since the incoming messages are all equal to $-1$, the update rule
of the WMS algorithm at the check node gives\[
\mu_{i\leftarrow j}^{(0)}=\left({\textstyle \underset{k\in N(j)\setminus i}{\prod}}\mbox{sgn}(\mu_{k\rightarrow j})\right)\min_{k'\in N(j)\setminus i}\left|\mu_{k'\rightarrow j}^{(0)}\right|=1,\]
for all $(i,j)\in\set{E}$. In the first iteration, the outgoing message
from the $i$th variable node to the $j$th check node is therefore\[
\mu_{i\rightarrow j}^{(1)}=\gamma_{i}+\beta{\textstyle \underset{k\in N(i)\setminus j}{\sum}}\mu_{i\leftarrow k}^{(0)}>-1+(d_{v}-1)\frac{2}{d_{v}-1}=1.\]
Moreover, one can show that $\mu_{i\rightarrow j}^{(\ell)}\rightarrow\infty$
as $\ell\rightarrow\infty$. Thus, the hard decision output is an
all-zeros codeword. Unfortunately, given this $\boldsymbol{\gamma}$,
we know that the ML output must be a nonzero codeword with maximal
Hamming weight. Therefore, WMS algorithm cannot provide an ML certificate
for $\beta>\frac{2}{d_{v}-1}$. One might worry that this effect may
be related to ties between ML codewords, but these can be avoided,
without affecting the above result, by adding a very small amount
of uniform random noise to the channel output LLRs.
\end{example}

\section{Convergence and Optimality Guarantees\label{sub:Convergence-and-Optimality}}

In this section, the optimality of codewords obtained by the AMP algorithms
and the WMS algorithms for LDPC codes is considered. We will show
that the AMP algorithm converges to a fixed point when the weight
factor $\beta(d_{v}-1)(d_{c}-1)<1$. Further, if there is a codeword
which satisfies the consistency conditions and uniquely maximizes
the converged value functions, it can be shown that the codeword is
the ML codeword. Similar to the analysis of the AMP decoding algorithm,
we first discuss the convergence of the WMS algorithm. Compared to
the convergence analysis of the AMP algorithm, a weaker condition
for the convergence of the WMS algorithm, $\beta(d_{v}-1)<1$, is
obtained. We also show that, if the converged messages satisfy the
consistency conditions, which are similar to the conditions for the
AMP algorithm, the optimality of the WMS codeword is guaranteed.

\subsection{Attenuated Max-product Decoding Algorithm}

Before showing that the AMP algorithm converges to a fixed point when
$\beta<\frac{1}{(d_{v}-1)(d_{c}-1)}$, we first introduce the following
tool lemma.
\begin{lem}
\label{lem: MaxDiffGeqDiffMax}For any two vectors $\vect{f},\,\vect{g}\in\mathbb{R}^{n}$,
the following inequality holds \begin{equation}
\max_{i}\left|f_{i}-g_{i}\right|\geq\left|\max_{i}f_{i}-\max_{i'}g_{i'}\right|.\label{eq: MaxDiffGeqDiffMax}\end{equation}
\end{lem}
\begin{IEEEproof}
See Appendix \ref{sec: PfOfMaxDiffGeqDiffMax}.\end{IEEEproof}
\begin{thm}
\label{thm: amp is a contraction}The operator $\op{A}$ is an $\left\Vert \cdot\right\Vert _{\infty}$
contraction on $\mathbb{R}^{2|\set{E}|}$ if \[
\beta<\frac{1}{(d_{v}-1)(d_{c}-1)}.\]
\end{thm}
\begin{IEEEproof}
Let $\boldsymbol{\mu},\boldsymbol{\nu}\in\mathbb{R}^{2|\set{E}|}$
be two vectors of AMP messages, and let $\boldsymbol{\mu}'=\op{A}[\boldsymbol{\mu}]$
and $\boldsymbol{\nu}'=\op{A}[\boldsymbol{\nu}]$. By the definition
of $\op{A}$ in (\ref{eq: DefOperatorA}) and the fact that for any
two vectors $\boldsymbol{f}$ and $\boldsymbol{g}$ over $\mathbb{R}$,
$|\sum_{i}(f_{i}-g_{i})|\leq\sum_{i}|f_{i}-g_{i}|$, $\left\Vert \boldsymbol{\mu}'-\boldsymbol{\nu}'\right\Vert _{\infty}$can
be upper bounded by\begin{align}
\left\Vert \boldsymbol{\mu}'-\boldsymbol{\nu}'\right\Vert _{\infty} & =\beta\max_{x\in\{0,1\},\,(i,j)\in\set{E}}\Bigg|\sum_{k\in N(i)\setminus j}\mu_{i\leftarrow k}(x)-\sum_{k'\in N(i)\setminus j}\nu_{i\leftarrow k'}(x)\Bigg|\nonumber \\
 & \leq\beta\max_{x\in\{0,1\},\,(i,j)\in\set{E}}\sum_{k\in N(i)\setminus j}\left|\mu_{i\leftarrow k}(x)-\nu_{i\leftarrow k}(x)\right|.\label{eq: nonnormal ineq 1}\end{align}
From (\ref{eq: AMPc2v}), the last term of the RHS in (\ref{eq: nonnormal ineq 1})
can be rewritten as \begin{align*}
\left|\mu_{i\leftarrow k}(x)-\nu_{i\leftarrow k}(x)\right| & =\Bigg|\max_{\vect{w}\in\set{S}_{k,i}(x)}\sum_{m\in N(k)\setminus i}\mu_{m\rightarrow k}(w_{m})-\max_{\vect{w}'\in\set{S}_{k,i}(x)}\sum_{m'\in N(k)\setminus i}\nu_{m'\rightarrow k}(w'_{m'})\Bigg|\\
 & \overset{\mbox{(a)}}{\leq}\max_{\vect{w}\in\set{S}_{k,i}(x)}\Bigg|\sum_{m\in N(k)\setminus i}\mu_{m\rightarrow k}(w_{m})-\sum_{m\in N(k)\setminus i}\nu_{m\rightarrow k}(w_{m})\Bigg|\\
 & \leq\max_{\vect{w}\in\set{S}_{k,i}(x)}\sum_{m\in N(k)\setminus i}\big|\mu_{m\rightarrow k}(w_{m})-\nu_{m\rightarrow k}(w_{m})\big|,\end{align*}
where the inequality (a) follows by Lemma \ref{lem: MaxDiffGeqDiffMax}.
Thus, the RHS in equation (\ref{eq: nonnormal ineq 1}) is upper bounded
by\begin{equation}
\beta\max_{x\in\{0,1\},\,(i,j)\in\set{E}}\sum_{k\in N(i)\setminus j}\max_{\vect{w}\in\mathcal{S}_{k,i}(x)}\sum_{m\in N(k)\setminus i}\left|\mu_{m\rightarrow k}\left(w_{m}\right)-\nu_{m\rightarrow k}\left(w_{m}\right)\right|.\label{eq: nonnormal ineq 2}\end{equation}
Since $\max(\vect f+\vect g)\leq\max\vect g+\max\vect f$, (\ref{eq: nonnormal ineq 2})
is further upper bounded by\begin{align*}
\lefteqn{\beta\sum_{\substack{k\in N(i)\setminus j\\
m\in N(k)\setminus i}
}\max_{\substack{x\in\{0,1\},\,(i,\, j)\in\set{E}\\
\vect{w}\in\mathcal{S}_{k,i}(x)}
}\left|\mu_{m\rightarrow k}\left(w_{m}\right)-\nu_{m\rightarrow k}\left(w_{m}\right)\right|}\\
 & \quad=\beta\left(d_{v}-1\right)\left(d_{c}-1\right)\max_{x\in\{0,1\},(i,j)\in\set{E}}\left|\mu_{i\rightarrow j}\left(x\right)-\nu_{i\rightarrow j}\left(x\right)\right|\\
 & \quad\overset{\mbox{(a)}}{<}\left\Vert \boldsymbol{\mu}-\boldsymbol{\nu}\right\Vert _{\infty},\end{align*}
where the inequality (a) follows from the fact that $\beta(d_{c}-1)(d_{v}-1)<1$.
This proves the theorem.
\end{IEEEproof}

\begin{rem}
Combining Theorem \ref{thm: amp is a contraction} with the contraction
mapping theorem shows that, for an arbitrary $(d_{v},d_{c})$-regular
LDPC code and any $0\leq\beta<\frac{1}{(d_{c}-1)(d_{v}-1)}$, the
AMP algorithm converges to a unique fixed point denoted by $\vect{\mu}^{*}$.
That is $\vect{\mu}^{(\ell)}\rightarrow\vect{\mu}^{*}$ as $\ell\rightarrow\infty$,
and $\vect{\mu}^{*}=\mathsf{A}[\vect{\mu}^{*}]$. We note that this
idea is very similar to the existence proof for optimal stationary
policies of discounted Markov decision processes. 
\end{rem}

For each $(i,j)\in\set{E}$, let $x_{i,j}^{*}\in\{0,1\}$ be the assignment
which uniquely maximizes $\mu_{i\rightarrow j}^{*}(x)$, and let $\vect{x}^{*}\in\{0,1\}^{n}$
be the vector returned by the AMP algorithm. For regular LDPC codes,
it suffices to show the ML optimality of $\vect{x}^{*}$ if the following
conditions hold.
\begin{defn}[AMP-consistency]
The assignment $\{x_{i,j}^{*}:(i,j)\in\set{E}\}$ is called \emph{AMP-consistent}
if $\vect{x}^{*}\in\set{C}$, $x_{i,j}^{*}=x_{i}^{*}$.%
{}\end{defn}
\begin{lem}
\label{lem: sum of mu vs corr} Consider a $(d_{v},d_{c})$-regular
LDPC code, and choose $\beta<\frac{1}{(d_{v}-1)(d_{c}-1)}$. For each
edge $(i,j)\in\set{E}$, let $\mu_{i\rightarrow j}^{*}(x)$ be the
fixed point, and let $x_{i,j}^{*}$ uniquely maximize $\mu_{i\rightarrow j}^{*}(x)$.
Then for any binary vector $\{x_{i,j}\}\in\{0,1\}^{|\set{E}|}$, \[
\sum_{(i,j)\in\set{E}}\mu_{i\rightarrow j}^{*}\left(x_{i,j}\right)\leq\sum_{(i,j)\in\set{E}}\left(1-x_{i,j}\right)\gamma_{i}+\beta\left(d_{v}-1\right)\left(d_{c}-1\right)\sum_{(i,j)\in\set{E}}\mu_{i\rightarrow j}^{*}\left(x_{i,j}^{*}\right),\]
with equality if and only if $\{x_{i,j}^{*}:\,(i,j)\in\set{E}\}$
is AMP-consistent, and $x_{i,j}=x_{i,j}^{*}$ for all $(i,j)\in\set{E}$. \end{lem}
\begin{IEEEproof}
See Appendix \ref{sec:PfOfSumOfMu}.
\end{IEEEproof}

\begin{rem}
From Lemma \ref{lem: sum of mu vs corr}, we know that when the assignment
$\{x_{i,j}^{*}\}$ is AMP-consistent, then \begin{align}
\sum_{(i,\, j)\in\set{E}}\mu_{i\rightarrow j}^{*}\left(x_{i,j}^{*}\right) & =\frac{d_{v}\sum_{i\in\set{V}_{L}}\left(1-x_{i}^{*}\right)\gamma_{i}}{1-\beta\left(d_{v}-1\right)\left(d_{c}-1\right)},\label{eq: mu_ijx* and corr}\end{align}
where $x_{i}^{*}=x_{i,j}^{*}$ for all $i\in\set{V}_{L}$ and $(i,j)\in\set{E}$.\end{rem}
\begin{thm}
Given the LLR vector $\boldsymbol{\gamma}\in\mathbb{R}^{n}$, let
the assignment $x_{i,j}^{*}$ uniquely maximize $\mu_{i\rightarrow j}^{*}(x)$
for all $(i,j)\in\set{E}$. If $\{x_{i,j}^{*}:\,(i,j)\in\set{E}\}$
is AMP-consistent, then $\boldsymbol{x}^{*}=\{x_{i}^{*}:\, i\in\set{V}_{L}\}$
is the ML codeword.\end{thm}
\begin{IEEEproof}
We prove that $\boldsymbol{x}^{*}$ is the ML codeword by showing
that $\boldsymbol{x}^{*}$ uniquely maximizes the correlation $\sum_{i\in\set{V}_{L}}(1-x_{i}^{*})\gamma_{i}$
over all codewords in $\mathcal{C}.$ 

Consider any codeword $\tilde{\boldsymbol{x}}\in\mathcal{C}$ such
that $\tilde{\boldsymbol{x}}\neq\boldsymbol{x}^{*}$, and $\{\tilde{x}_{i,j}\}\in\{0,1\}^{|\set{E}|}$
be the corresponding binary vector with $\tilde{x}_{i,j}=\tilde{x}_{i}$
for all $j\in N(i)$. From (\ref{eq: amp mu_ij update}), we know\begin{align}
\sum_{(i,\, j)\in\set{E}}\left(1-\tilde{x}_{i,j}\right)\gamma_{i} & =\sum_{(i,\, j)\in\set{E}}\mu_{i\rightarrow j}^{*}\left(\tilde{x}_{i,j}\right)-\beta\sum_{\substack{(i,\, j)\in\set{E}\\
k\in N(i)\setminus j}
}\max_{\boldsymbol{w}\in\mathcal{S}_{k,i}(\tilde{x}_{i,j})}\sum_{m\in N(k)\setminus i}\mu_{m\rightarrow k}^{*}\left(w_{m}\right).\label{eq: AMP corr xtilde}\end{align}
By the fact that $\tilde{\boldsymbol{x}}$ is also in $\mathcal{S}_{k,i}(\tilde{x}_{i,j})$,
we have\[
\max_{\boldsymbol{w}\in\mathcal{S}_{k,i}(\tilde{x}_{i,j})}\sum_{m\in N(k)\setminus i}\mu_{m\rightarrow k}^{*}\left(w_{m}\right)\geq\sum_{m\in N(k)\setminus i}\mu_{m\rightarrow k}^{*}\left(\tilde{x}_{m,k}\right).\]
Therefore, the RHS in (\ref{eq: AMP corr xtilde}) is upper bounded
by\begin{align}
\lefteqn{\sum_{(i,j)\in\set{E}}\mu_{i\rightarrow j}^{*}\left(\tilde{x}_{i,j}\right)-\beta\sum_{\substack{(i,j)\in\set{E},\, k\in N(i)\setminus j\\
m\in N(k)\setminus i}
}\mu_{m\rightarrow k}^{*}\left(\tilde{x}_{m,k}\right)}\nonumber \\
 & \qquad=\left(1-\beta\left(d_{v}-1\right)\left(d_{c}-1\right)\right)\sum_{(i,j)\in\set{E}}\mu_{i\rightarrow j}^{*}\left(\tilde{x}_{i,j}\right).\label{eq: hat x_ij corr}\end{align}
Since $x_{i,j}^{*}$ uniquely maximizes $\sum_{(i,\, j)\in\set{E}}\mu_{i\rightarrow j}^{*}\left(x\right)$,
the RHS in (\ref{eq: hat x_ij corr}) is less than\[
\left(1-\beta\left(d_{v}-1\right)\left(d_{c}-1\right)\right)\sum_{(i,\, j)\in\set{E}}\mu_{i\rightarrow j}^{*}\left(x_{i,j}^{*}\right).\]
Thus, we have \begin{align*}
\sum_{i\in\set{V}_{L}}\left(1-\tilde{x}_{i}\right)\gamma_{i} & <\frac{1}{d_{v}}\left(1-\beta\left(d_{v}-1\right)\left(d_{c}-1\right)\right)\sum_{(i,\, j)\in\set{E}}\mu_{i\rightarrow j}^{*}\left(x_{i,j}^{*}\right)\\
 & \overset{\mbox{(a)}}{=}\sum_{i\in\set{V}_{L}}\left(1-x_{i}^{*}\right)\gamma_{i},\end{align*}
where (a) follows from (\ref{eq: mu_ijx* and corr}). This shows that
$\boldsymbol{x}^{*}$ uniquely maximizes the correlation $\sum_{i\in\set{V}_{L}}\left(1-x_{i}\right)\gamma_{i}$
over all $\boldsymbol{x}\in\mathcal{C}$ and is therefore the ML codeword.
\end{IEEEproof}

\subsection{Weighted Min-sum Decoding Algorithm\label{sub:discountedWMS}}

Before showing the optimality of the WMS algorithm, we first introduce
a consistency condition for WMS decoding.
\begin{defn}[WMS-consistency]
 Let $\mu_{i\rightarrow j}^{(\ell)}$ be the message passed from
the $i$th bit to the $j$th check in the $\ell$th iteration, and
$\mu_{i\leftarrow j}^{(\ell)}$ be the message passed from $j$th
check to $i$th bit, defined in (\ref{eq: WMS-c2v}). The message
vector $\vect{\mu}^{(\ell)}$ is called \emph{WMS-consistent} if,
for each bit $i\in\set{V}_{L}$, it satisfies 1) $\sgn(\mu_{i\rightarrow j}^{(\ell)})=\sgn(\mu_{i\rightarrow j'}^{(\ell)})$
for $j,j'\in N(i)$, 2) $\sgn(\mu_{i\leftarrow j}^{(\ell)})=\sgn(\mu_{i\rightarrow j}^{(\ell)})$
for $j\in N(i)$, and 3) $\sgn(\gamma_{i}+\beta\smash{\sum_{j\in N(i)}\mu_{i\leftarrow j}^{(\ell)}})=\sgn(\mu_{i\rightarrow j'}^{(\ell)})$
for $j'\in N(i).$
\end{defn}

When the WMS messages satisfy the WMS-consistency conditions, the
following theorem shows that the corresponding hard decisions return
a codeword.
\begin{thm}
\label{thm: xhat is a cw} If the WMS messages in the $\ell$th iteration
are WMS-consistent, then the hard decisions \[
\hat{x}_{i}=\frac{1}{2}\left(1-\sgn\left(\gamma_{i}+\beta\sum_{j\in N(i)}\mu_{i\leftarrow j}^{(\ell)}\right)\right)\]
for $i=1,\dots,n$ give a codeword.\end{thm}
\begin{IEEEproof}
We prove this result by contradiction. Assume that $\hat{\boldsymbol{x}}$
is not a codeword. There exists at least one unsatisfied parity check
node. Let $j\in\set{V}_{R}$ be the unsatisfied parity check node
and $N(j)$ be the neighbors of $j$. Since $\sum_{i\in N(j)}\hat{x}_{i}=1\mod2,$
we have\begin{align*}
-1 & =\prod_{i\in N(j)}\sgn\left(\gamma_{i}+\beta\sum_{j'\in N(i)}\mu_{i\leftarrow j'}^{(\ell)}\right)\\
 & =\prod_{i\in N(j)}\sgn\left(\mu_{i\rightarrow j}^{(\ell)}\right).\end{align*}
 Consider the message passed from the $j$th check to the $i$th bit.
From the WMS update rule,\begin{align*}
\mu_{i\leftarrow j}^{(\ell)} & =\left(\prod_{m\in N(j)\setminus i}\sgn\left(\mu_{m\rightarrow j}^{(\ell)}\right)\right)\times\min_{m\in N(j)\setminus i}\left|\mu_{m\rightarrow j}^{(\ell)}\right|\\
 & =-\sgn\left(\mu_{i\rightarrow j}^{(\ell)}\right)\times\min_{m\in N(j)\setminus i}\left|\mu_{m\rightarrow j}^{(\ell)}\right|.\end{align*}
This contradicts the condition 2) of WMS-consistency.
\end{IEEEproof}

Next, we consider the optimality of the solution returned by the WMS
decoder. Similar to the analysis of the AMP algorithm, we first discuss
the convergence of the WMS messages. When the WMS messages converge
to a fixed point, we show that the corresponding hard decisions give
an optimal codeword if the fixed point is WMS-consistent.

To show the convergence of the WMS algorithm, we first introduce the
following lemma.
\begin{lem}
\label{lem:ContractionLemma} Consider two WMS message vectors $\boldsymbol{\mu},\boldsymbol{\nu}\in\mathbb{R}^{|\set{E}|}$.
Let $i\in\set{V}_{L}$, $k\in\set{V}_{R}$ and $(i,\, k)\in\set{E}$,
and define\begin{align*}
d_{i,k} & \triangleq\left|\left(\prod_{m\in N(k)\setminus i}\sgn\left(\mu_{m\rightarrow k}\right)\right)\min_{m'\in N(k)\setminus i}\left|\mu_{m'\rightarrow k}\right|\right.\\
 & \qquad\left.-\left(\prod_{m\in N(k)\setminus i}\sgn\left(\nu_{m\rightarrow k}\right)\right)\min_{m'\in N(k)\backslash i}\left|\nu_{m'\rightarrow k}\right|\right|.\end{align*}
Then, \[
\max_{m\in N(k)\backslash i}\left|\mu_{m\rightarrow k}-\nu_{m\rightarrow k}\right|\geq d_{i,k}.\]
\end{lem}
\begin{IEEEproof}
See Appendix \ref{sec:Proof-of-ContractionLemma}.
\end{IEEEproof}

To show the convergence of the WMS messages, it will suffice to show
that the WMS operator $\op{W}$ is an $\left\Vert \cdot\right\Vert _{\infty}$
contraction. The following theorem provides a precise statement.
\begin{thm}
\label{thm: LLR converge}For all LLR vectors and message vectors,
the WMS operator $\op{W}$ is an $\left\Vert \cdot\right\Vert _{\infty}$
contraction if \[
\beta\left(d_{v}-1\right)<1.\]
\end{thm}
\begin{IEEEproof}
Using Lemma \ref{lem:ContractionLemma}, one can upper bound $\left\Vert \op{W}[\boldsymbol{\mu}]-\op{W}[\boldsymbol{\nu}]\right\Vert _{\infty}$
in a straightforward manner to get\begin{align*}
\left\Vert \op{W}\left[\boldsymbol{\mu}\right]-\op{W}\left[\boldsymbol{\nu}\right]\right\Vert _{\infty} & \leq\max_{(i,\, j)\in\set{E}}\beta{\textstyle \underset{k\in N(i)\backslash j}{\sum}}d_{i,k}\\
 & \leq\beta\left(d_{v}-1\right)\max_{(i,\, k)\in\set{E}}d_{i,k}\\
 & \leq\beta\left(d_{v}-1\right)\max_{\substack{(i,\, k)\in\set{E}\\
m\in N(k)\backslash i}
}\left|\mu_{m\rightarrow k}-\nu_{m\rightarrow k}\right|\\
 & =\beta\left(d_{v}-1\right)\left\Vert \boldsymbol{\mu}-\boldsymbol{\nu}\right\Vert _{\infty}.\end{align*}
This implies that $\op{W}$ is a $\left\Vert \cdot\right\Vert _{\infty}$
contraction. \end{IEEEproof}
\begin{rem}
Combining this with the contraction mapping theorem shows that, for
an arbitrary $(d_{v},d_{c})$-regular LDPC code and any $0\leq\beta<\frac{1}{d_{v}-1}$,
the WMS algorithm converges to a unique fixed point, $\mu_{i\rightarrow j}^{(\ell)}\rightarrow\mu_{i\rightarrow j}^{*}$
and $\mu_{i\leftarrow j}^{(\ell)}\rightarrow\mu_{i\leftarrow j}^{*}$,
as the number of iterations goes to infinity. 
\end{rem}

For any WMS-consistent fixed point, there are two ways to prove the
optimality of the hard decision output. One way is by looking at \emph{Problem-P
}directly, which has been shown in our earlier work \cite{Jian-istc10}.
We generalize the definition of \emph{minimal $T$-local deviation}
in \cite{Arora-stoc09} to $T\geq\frac{1}{4}\mbox{girth}(\set{G})$.
By using the generalized \emph{minimal $T$-local deviation}, it can
be shown that, if the fixed point is WMS-consistent, the corresponding
hard decision bits also return a locally optimal codeword. By the
fact that local optimality implies global optimal and LP optimal,
the hard decision is an LP and ML codeword. A summary of \cite{Jian-istc10}
is provided in Appendix \ref{sec: Extension of ADS}. 

The other method, which is introduced in the rest of this section,
is by examining the optimality in \emph{Problem-D}. We construct a
dual witness according to the method introduced in \cite{Koetter-ita06}.
The following lemma shows that the vector $\boldsymbol{\tau}^{*}\in\mathbb{R}^{|\set{E}|}$,
which is constructed from the fixed-point messages $\mu_{i\rightarrow j}^{*}$
and $\mu_{i\leftarrow j}^{*}$, is a dual feasible point of \emph{Problem-P}. 
\begin{lem}
\label{lem: dual feasible constrcut} Consider the WMS algorithm with
$\beta<\frac{1}{d_{v}-1}$ over a $(d_{v},d_{c})$-regular LDPC code.
The vector $\boldsymbol{\tau}^{*}\in\mathbb{R}^{|\set{E}|}$ defined
by \begin{align}
\tau_{i,j}^{*} & =\frac{1}{d_{v}}\left(\mu_{i\rightarrow j}^{*}-\beta\left(d_{v}-1\right)\mu_{i\leftarrow j}^{*}\right)\label{eq: tau in terms of mu}\end{align}
is a dual feasible point of \emph{Problem-P}.\end{lem}
\begin{IEEEproof}
Fix a variable node $i\in\set{V}_{L}$. The sum of the dual variables
on the edges incident to $i$ is given by \begin{align*}
\sum_{j\in N(i)}\tau_{i,j}^{*} & =\frac{1}{d_{v}}\sum_{j\in N(i)}\left(\mu_{i\rightarrow j}^{*}-\beta\left(d_{v}-1\right)\mu_{i\leftarrow j}^{*}\right)\\
 & =\frac{1}{d_{v}}\sum_{j\in N(i)}\left(\mu_{i\rightarrow j}^{*}-\sum_{k\in N(i)\setminus j}\beta\mu_{i\leftarrow k}^{*}\right)\\
 & =\gamma_{i}.\end{align*}
This proves the lemma.\end{IEEEproof}
\begin{rem}
Compared to the construction in \cite{Koetter-ita06}, Lemma \ref{lem: dual feasible constrcut}
is a simplified version by just considering one-step update of the
WMS messages. In \cite{Koetter-ita06}, min-sum messages over $L$
iterations are considered. For a computation tree $\mathcal{T}_{j}^{2L}$
of depth $2L$ rooted at check node $j$, those min-sum messages are
used to generate an assignment $\boldsymbol{\tau}(j,L)$ to edges
in $\mathcal{T}_{j}^{2L}$. Koetter and Vontobel showed that the dual
feasible point $\boldsymbol{\tau}^{*}$ can be obtained by averaging
$\boldsymbol{\tau}(j,L)$ over all $j\in\set{V}_{R}$. Since the number
of leaf nodes in a computation tree increases doubly exponentially,
a weight factor $\alpha$ is introduced to attenuate the influence
of the leaves of the computation tree. In our analysis, by the fact
that the WMS messages satisfy a fixed-point equation, we simplify
the construction and consider only the assignments on the top level
of computation tree. Next, we will show that the proposed dual-feasible
point $\boldsymbol{\tau}^{*}$ is also a dual-optimal point in \emph{Problem-D}
if it is constructed from a WMS-consistent fixed point.
\end{rem}

For a $j\in\set{V}_{R}$, let $\boldsymbol{\tau}_{j}^{*}\in\mathbb{R}^{d_{c}}$
denote the assignments on the edges incident to $j$, $\{\tau_{i,j}^{*}:i\in N(j)\}$,
and let $\boldsymbol{\mu}_{j}^{*}=\{\mu_{i\rightarrow j}^{*}:\, i\in N\left(j\right)\}$
be the set of messages to $j$. Without loss of generality, we can
sort the vertices in $N(j)$ by $\left(\smash{i_{1},\, i_{2},\dots,\, i_{d_{c}}}\right)$
such that $\left|\smash{\mu_{i_{1}\rightarrow j}^{*}}\right|\leq\left|\smash{\mu_{i_{2}\rightarrow j}^{*}}\right|\leq\dots\leq\left|\smash{\mu_{i_{d_{c}}\rightarrow j}^{*}}\right|$.
With this order, $\boldsymbol{\tau}_{j}^{*}$ is rearranged into a
vector $\boldsymbol{t}\in\mathbb{R}^{d_{c}}$, where $t_{k}=\tau_{i_{k},j}^{*}$
for $k=1,2,\dots,d_{c}$. Also, we define two vectors $\overset{\rightarrow}{\boldsymbol{\mu}},\overset{\leftarrow}{\boldsymbol{\mu}}\in\mathbb{R}^{d_{c}}$
with $\overset{\rightarrow}{\mu}_{k}=\mu_{i_{k}\rightarrow j}^{*}$
and $\overset{\leftarrow}{\mu}_{k}=\mu_{i_{k}\leftarrow j}^{*}$ for
$k=1,2,\dots,d_{c}$, respectively. Given a vector $\vect{f}$, we
use $\sgn(\vect{f})$ to denote a vector which is composed of the
sign of each entry in $\vect{f}$. Finally, we use $\boldsymbol{1}$
to represent an all-one vector, and the dimension is determined in
the context of equations.

The following lemma shows that an affine function of $\sgn(\vect{\mu}_{j}^{*})$
minimizes the inner product $\left\langle \boldsymbol{w},\boldsymbol{t}\right\rangle $
for all $\boldsymbol{w}\in\set{L}$ when the fixed point is WMS-consistent.
Recall that $\set{L}$ is defined in (\ref{eq: ListOfLocalCodeword}).
\begin{lem}
\label{lem: hd minimizes sum}Consider a $(d_{v},d_{c})$-regular
LDPC code. For some $j\in\set{V}_{R}$, if the WMS algorithm with
$\beta<\frac{1}{d_{v}-1}$ converges to a WMS-consistent fixed point
$\mu_{i\rightarrow j}^{*}$ and $\mu_{i\leftarrow j}^{*}$ for all
$(i,j)\in\set{E}$, then\begin{align}
\argmin_{\boldsymbol{w}\in\set{L}}\left\langle \boldsymbol{w},\boldsymbol{\tau}_{j}^{*}\right\rangle  & =\frac{1}{2}\left(\vect{1}-\sgn\left(\boldsymbol{\mu}_{j}^{*}\right)\right).\label{eq: local min}\end{align}
\end{lem}
\begin{IEEEproof}
Since messages are WMS-consistent, we know that the RHS in (\ref{eq: local min})
satisfies the $j$th check node from Theorem \ref{thm: xhat is a cw}.
From (\ref{eq: tau in terms of mu}), the LHS in (\ref{eq: local min})
can be rewritten as\begin{align}
\left\langle \boldsymbol{w},\boldsymbol{\tau}_{j}^{*}\right\rangle  & =\sum_{k=1}^{d_{c}}w_{k}t_{k}\nonumber \\
 & =\frac{1}{d_{v}}\sum_{k=1}^{d_{c}}w_{k}\left(\overset{\rightarrow}{\mu}_{k}-\beta\left(d_{v}-1\right)\overset{\leftarrow}{\mu}_{k}\right)\nonumber \\
 & \overset{\mbox{(a)}}{=}\frac{1}{d_{v}}\sum_{k=1}^{d_{c}}w_{k}\sgn\left(\overset{\rightarrow}{\mu}_{k}\right)\left(\big|\overset{\rightarrow}{\mu}_{k}\big|-\beta\left(d_{v}-1\right)\big|\overset{\leftarrow}{\mu}_{k}\big|\right),\label{eq: local inner prod}\end{align}
where the equality (a) holds by condition 2) of WMS-consistency. From
the update rule of the WMS algorithm, one can show that \[
\big|\overset{\leftarrow}{\mu}_{k}\big|=\begin{cases}
\big|\overset{\rightarrow}{\mu}_{2}\big|, & \mbox{ when }k=1,\\
\big|\overset{\rightarrow}{\mu}_{1}\big|, & \mbox{ otherwise.}\end{cases}\]
Thus, the summation in (\ref{eq: local inner prod}) becomes\begin{align*}
\lefteqn{\frac{1}{d_{v}}w_{1}\sgn\left(\overset{\rightarrow}{\mu}_{1}\right)\left(\big|\overset{\rightarrow}{\mu}_{1}\big|-\beta\left(d_{v}-1\right)\big|\overset{\rightarrow}{\mu}_{2}\big|\right)}\\
 & \qquad+\frac{1}{d_{v}}\sum_{k=2}^{d_{c}}w_{k}\sgn\left(\overset{\rightarrow}{\mu}_{k}\right)\left(\big|\overset{\rightarrow}{\mu}_{k}\big|-\beta\left(d_{v}-1\right)\big|\overset{\rightarrow}{\mu}_{1}\big|\right).\end{align*}
Since $0\leq\beta\left(d_{v}-1\right)<1$, one can show that\begin{align*}
\left|\big|\overset{\rightarrow}{\mu}_{1}\big|-\beta\left(d_{v}-1\right)\big|\overset{\rightarrow}{\mu}_{2}\big|\right| & \leq\left|\big|\overset{\rightarrow}{\mu}_{k}\big|-\beta\left(d_{v}-1\right)\big|\overset{\rightarrow}{\mu}_{1}\big|\right|\end{align*}
for $k\geq2$. Thus, the minimum is achieved by choosing \begin{align*}
w_{k} & =\begin{cases}
{\displaystyle \frac{1}{2}\left(1-\sgn\left(\overset{\rightarrow}{\mu}_{k}\right)\right)} & \mbox{ for }k=2,3,\dots,d_{c},\\
{\displaystyle \sum_{m=2}^{d_{c}}w_{m}}\mod2 & \mbox{ for }k=1.\end{cases}\end{align*}
By the fact that $\frac{1}{2}(\vect{1}-\sgn(\boldsymbol{\mu}_{j}^{*}))$
satisfies the $j$th check node, thus $w_{1}=\frac{1}{2}(1-\sgn(\overset{\rightarrow}{\mu}_{1}))=\frac{1}{2}(1-\sgn(\mu_{i_{1}\rightarrow j}^{*}))$.
This completes the proof.
\end{IEEEproof}

\begin{rem}
The proof of Lemma \ref{lem: hd minimizes sum} employs part of the
observation in the proof of \cite[Lemma 3]{Koetter-ita06}. Given
a check node, the absolute values of all but one outgoing WMS messages
are the same. The only different absolute value of the outgoing message
will be passed along the edge that the smallest absolute value of
incoming message was passed on. With this observation, we know that
the corresponding binary value $w_{1}$ will depend on the other binary
values $w_{2},\dots,w_{d_{c}}$. Since min-sum messages are not guaranteed
to converge, Koetter and Vontobel computed the dual feasible point
using computation trees of depth greater than one. In order to offset
the influence of the exponential weighting of the messages from the
leaf nodes, a large initial value assumption is required. With this
large initial value assumption, they showed that the constructed dual
feasible point is an optimal point in \emph{Problem-D}.
\end{rem}

Let $\boldsymbol{\tau}\in\mathbb{R}^{|\set{E}|}$ and $g(\boldsymbol{\tau})$
be the objective function in \emph{Problem-D}. Let $\boldsymbol{w}^{(j)}\triangleq\frac{1}{2}(\vect{1}-\sgn(\boldsymbol{\mu}_{j}^{*}))$
be the local assignment to check $j$. By Lemma \ref{lem: dual feasible constrcut}
and Lemma \ref{lem: hd minimizes sum}, one can show that the optimal
value of the objective function in \emph{Problem-D} given $\boldsymbol{\tau}^{*}$
is \[
g(\boldsymbol{\tau}^{*})=\sum_{j\in\set{V}_{R}}\left\langle \boldsymbol{\tau}_{j}^{*},\boldsymbol{w}^{(j)}\right\rangle .\]
To find the optimal solution of \emph{Problem-D}, one needs to search
over all $\boldsymbol{\tau}$ in the dual-feasible set and find the
maximum of $g(\boldsymbol{\tau})$. Let the optimal value of \emph{Problem-P}
and the optimal value of \emph{Problem-D} be $f^{*}$ and $g^{*}$,
respectively. Since $\vect{\tau}^{*}$ is in the feasible set of \emph{Problem-D},
it is obvious that $g^{*}\geq g(\boldsymbol{\tau}^{*})$. In the following
theorem, we show that if the fixed point $\vect{\mu}^{*}\triangleq\{\mu_{i\rightarrow j}^{*}:(i,j)\in\set{E}\}$
is WMS-consistent, the proposed dual-feasible point $\boldsymbol{\tau}^{*}$
actually achieves the maximum, that is, $g^{*}=g(\boldsymbol{\tau}^{*})$.
Also, the corresponding hard decisions return an optimal codeword,
i.e., an ML codeword.
\begin{thm}
\label{thm: x is ML}Consider the WMS algorithm with $\beta<\frac{1}{d_{v}-1}$.
If the message vector $\vect{\mu}^{(\ell)}$ converges to a WMS-consistent
fixed point, $\vect{\mu}^{*}$, then the hard decision bits $\boldsymbol{x}^{*}\in\{0,1\}^{n}$
with \[
x_{i}^{*}=\frac{1}{2}\left(\vect{1}-\sgn\left(\gamma_{i}+\beta\sum_{j\in N(i)}\mu_{i\leftarrow j}^{*}\right)\right)\]
is a codeword. Also, $\boldsymbol{x}^{*}$ is LP optimal and, hence,
ML optimal.\end{thm}
\begin{IEEEproof}
Let $\boldsymbol{\tau}^{*}$ be a dual feasible point constructed
as proposed in Lemma \ref{lem: dual feasible constrcut}. Let $\boldsymbol{w}^{(j)}\in\set{L}$
be the binary vector that minimizes the inner product $\left\langle \smash{\boldsymbol{w},\boldsymbol{\tau}_{j}^{*}}\right\rangle $
over all $\boldsymbol{w}\in\set{L}$. Then, from Lemma \ref{lem: hd minimizes sum},
we know $\boldsymbol{w}^{(j)}=\frac{1}{2}\left(\vect{1}-\sgn\left(\smash{\boldsymbol{\mu}_{j}^{*}}\right)\right)$
for each $j\in\set{V}_{R}$. Since the fixed point $\vect{\mu}^{*}$
is WMS-consistent, by Theorem \ref{thm: xhat is a cw}, it can be
shown that $\boldsymbol{w}^{(j)}=\left\{ x_{i}^{*}:i\in N\left(j\right)\right\} $,
where $x_{i}^{*}=\frac{1}{2}\left(1-\sgn\left(\smash{\gamma_{i}+\beta\sum_{j\in N(i)}\mu_{i\leftarrow j}^{*}}\right)\right)$
is the hard decision of the $i$th bit. 

In the following proof, we will show that $\boldsymbol{x}^{*}$ is
LP optimal by contradiction. Assume that $\boldsymbol{x}^{*}$ does
not minimize \emph{Problem-P}, then we have\begin{align*}
f^{*} & <\sum_{i\in\set{V}_{L}}\gamma_{i}x_{i}^{*}\overset{\mbox{(a)}}{=}\sum_{i\in\set{V}_{L}}\left(\sum_{j\in N\left(i\right)}\tau_{i,j}^{*}\right)x_{i}^{*}\\
 & \overset{\mbox{(b)}}{=}\sum_{j\in\set{V}_{R}}\left(\sum_{i\in N\left(j\right)}\tau_{i,j}^{*}x_{i}^{*}\right)=\sum_{j\in\set{V}_{R}}\left\langle \boldsymbol{\tau}_{j}^{*},\boldsymbol{w}^{(j)}\right\rangle =g\left(\boldsymbol{\tau}^{*}\right)\leq g^{*},\end{align*}
where (a) follows from Lemma \ref{lem: dual feasible constrcut},
and (b) is a result of the WMS-consistency conditions. But, weak duality
implies that $f^{*}\geq g^{*}$, and this gives a contradiction. Thus,
$\boldsymbol{x}^{*}$ minimizes the primal problem, and hence, is
LP optimal. Moreover, since $\boldsymbol{x}^{*}\in\mathcal{C}$, it
is also an ML codeword. \end{IEEEproof}
\begin{rem}
Consider the WMS algorithm on a $(d_{v},d_{c})$-regular LDPC code
with $\beta<\frac{1}{d_{v}-1}$. From Theorem \ref{thm: x is ML},
we are able to check the optimality of the WMS solution by testing
the WMS-consistency conditions . If the messages satisfy the consistency
conditions, then the hard decision bits return an ML codeword. 
\end{rem}

\section{Weighted Min-sum Decoding with $\beta=\frac{1}{d_{v}-1}$\label{sec: BetaEq1/(dv-1)}}

We first introduce some notation and definitions. We denote the WMS
messages $\{\mu_{i\rightarrow j}^{(\ell)}:\,(i,j)\in\set{E}\}$ with
$\beta=\frac{1}{d_{v}-1}$ in the $\ell$th iteration by a vector
$\boldsymbol{\mu}^{(\ell)}\in\mathbb{R}^{|\set{E}|}$. The hard decisions
computed by $\boldsymbol{\mu}^{(L_{0})}$ are denoted by a binary
vector $\boldsymbol{x}^{(L_{0})}\in\{0,1\}^{n}$. For the WMS algorithm
with $\beta=\frac{\delta}{d_{v}-1}$ and $0\leq\delta<1$, we use
the vectors $\boldsymbol{\mu}_{\delta}^{(\ell)}\in\mathbb{R}^{|\set{E}|}$
and $\boldsymbol{\mu}_{\delta}^{*}\in\mathbb{R}^{|\set{E}|}$ to denote
the messages in the $\ell$th iteration and the fixed-point messages,
respectively. The collection of hard decision bits computed using
$\boldsymbol{\mu}_{\delta}^{*}$ is denoted by a vector $\boldsymbol{x}_{\delta}^{*}\in\{0,1\}^{n}$.
Moreover, for any WMS message vector $\boldsymbol{\mu}\in\mathbb{R}^{|\set{E}|}$,
the vector $|\boldsymbol{\mu}|\in\mathbb{R}_{+}^{|\set{E}|}$ consists
of the absolute value of each element of $\boldsymbol{\mu}$. For
any two WMS message vectors $\boldsymbol{\mu},\boldsymbol{\nu}\in\mathbb{R}^{|\set{E}|}$,
we use $\boldsymbol{\mu}\overset{s}{=}\boldsymbol{\nu}$ to denote
that $\sgn(\mu_{i\rightarrow j})=\sgn(\nu_{i\rightarrow j})$ for
all $(i,j)\in\set{E}$. When comparing two vectors, we use the partial
order $\boldsymbol{\mu}\succ\boldsymbol{\nu}$ to denote $\mu_{i\rightarrow j}>\nu_{i\rightarrow j}$
for all $(i,j)\in\set{E}$, and $\boldsymbol{\mu}\succeq\boldsymbol{\nu}$
to denote $\mu_{i\rightarrow j}\geq\nu_{i\rightarrow j}$ for all
$(i,j)\in\set{E}$. In the sequel, $\{\boldsymbol{\mu}\}$ and $\{\boldsymbol{\mu}_{\delta}\}$
denote sequences of WMS message vectors $\{\boldsymbol{\mu}^{(\ell)}:\ell=1,2,\dots\}$
and $\{\boldsymbol{\mu}_{\delta}^{(\ell)}:\ell=1,2,\dots\}$, respectively.
We extend the definition of the WMS operator in (\ref{eq: WMSOperator})
to $\op{W}_{\delta}$ for $\beta=\frac{\delta}{d_{v}-1}$. The conditions
for the operator $\op{W}_{\delta}$ to preserve the partial order
of the absolute value of the WMS messages are introduced in the following
lemma.
\begin{lem}
\label{lem: OperatorIsAbsMonotone}Consider a $(d_{v},d_{c})$-regular
LDPC code and a particular LLR vector $\boldsymbol{\gamma}\in\mathbb{R}^{n}$.
Let $\boldsymbol{\mu},\boldsymbol{\nu}\in\mathbb{R}^{|\set{E}|}$
be two WMS-consistent message vectors. If $\boldsymbol{\mu}\overset{s}{=}\boldsymbol{\nu}$
and $|\boldsymbol{\mu}|\succeq|\boldsymbol{\nu}|\succeq\frac{\|\boldsymbol{\gamma}\|_{\infty}}{\delta}\boldsymbol{1}$,
then $|\op{W}_{\delta}[\boldsymbol{\mu}]|\succeq|\op{W}_{\delta}[\boldsymbol{\nu}]|$
and $\op{W}_{\delta}[\boldsymbol{\mu}]\overset{s}{=}\op{W}_{\delta}[\boldsymbol{\nu}]\overset{s}{=}\boldsymbol{\mu}$.\end{lem}
\begin{IEEEproof}
See Appendix \ref{sec:pfOperatorIsAbsMonotone}
\end{IEEEproof}

%
{}

When $\beta=\frac{1}{d_{v}-1}$, one may observe three kinds of trajectories
of the WMS messages. They can converge to a fixed point, oscillate,
or diverge to $\pm\infty$. In this section, we are interested in
the case when the sequence of WMS message vectors, $\{\boldsymbol{\mu}\}$,
is divergent and WMS-consistent. We formalize this case by the following
definition. 
\begin{defn}
\label{def: DivergeConsistent}A sequence of WMS message vectors,
$\{\boldsymbol{\mu}\}$, is \emph{divergent and consistent }if 1)
for all $(i,j)\in\mathcal{E}$, the absolute value of the WMS message,
$|\mu_{i\rightarrow j}^{(\ell)}|$, goes to infinity, and 2) there
exists an integer $L>0$ such that $\boldsymbol{\mu}^{(\ell)}$ is
WMS-consistent whenever $\ell\geq L$.
\end{defn}

%
{}

Given two positive integers $L_{1}>L_{0}$, to simplify notation,
we denote $I=\{L_{0},L_{0}+1,\dots,L_{1}\}$ by $I=[L_{0},L_{1}]$
when it is clear from context that $I$ contains integers. A property
of the sequence of WMS message vectors, $\{\vect{\mu}_{\delta}\}$,
is introduced in the following definition. 
\begin{defn}[Block-wise monotone property]
\label{def: BMPI}A sequence of WMS message vectors, $\{\boldsymbol{\mu}_{\delta}\}$,
is said to have \emph{block-wise monotone property} in interval $I=[L_{0},\, L_{1}]$
denoted by BMP($I$), if for all $\ell\in I$, 1) $\boldsymbol{\mu}_{\delta}^{(\ell)}$
is WMS-consistent, 2) $\boldsymbol{\mu}_{\delta}^{(\ell)}\overset{s}{=}\boldsymbol{\mu}_{\delta}^{(L_{0})}$,
3) $|\boldsymbol{\mu}_{\delta}^{(\ell)}|\succeq\frac{\|\boldsymbol{\gamma}\|_{\infty}}{\delta}\boldsymbol{1}$,
and 4) $|\boldsymbol{\mu}_{\delta}^{(L_{1})}|\succeq|\boldsymbol{\mu}_{\delta}^{(L_{0})}|$. 
\end{defn}

In the following analysis, we show that, if there is an interval $I_{0}=[L_{0},\, L_{1}]$
such that the sequence of WMS message vectors, $\{\boldsymbol{\mu}_{\delta}\}$,
satisfies BMP($I_{0}$), then $\{\boldsymbol{\mu}_{\delta}\}$ also
satisfies BMP($I_{k}$) for all intervals $I_{k}=[L_{0}+k(L_{1}-L_{0}),\, L_{1}+k(L_{1}-L_{0})]$.
We first show that if $\{\boldsymbol{\mu}_{\delta}\}$ satisfies BMP($I_{0}$),
then $\{\boldsymbol{\mu}_{\delta}\}$ also satisfies BMP($I_{1}$),
where $I_{1}=[L_{1},\, L_{1}+(L_{1}-L_{0})]$.
\begin{lem}
\label{lem: ExtendOneBlock}Let $\boldsymbol{\gamma}\in\mathbb{R}^{n}$
be the received LLRs, and consider the sequence of WMS message vectors
$\{\boldsymbol{\mu}_{\delta}\}$ of a $(d_{v},d_{c})$-regular LDPC
code. Suppose there exists an interval $I_{0}=[L_{0},\, L_{1}]$ such
that $\{\boldsymbol{\mu}_{\delta}\}$ satisfies BMP($I_{0}$), then
\begin{equation}
\boldsymbol{\mu}_{\delta}^{(\ell'+L_{1})}\overset{s}{=}\boldsymbol{\mu}_{\delta}^{(\ell'+L_{0})}\label{eq: SameSign}\end{equation}
and \begin{equation}
\left|\boldsymbol{\mu}_{\delta}^{(\ell'+L_{1})}\right|\succeq\left|\boldsymbol{\mu}_{\delta}^{(\ell'+L_{0})}\right|\label{eq: PartialOrder}\end{equation}
for all $\ell'=0,\,1,\,\dots,\, L_{1}-L_{0}$. \end{lem}
\begin{IEEEproof}
We prove this lemma by induction. The base case, $\ell'=0$, is obtained
since conditions 2) and 4) of BMP($I_{0}$) are satisfied. %
{}

For the inductive step, suppose that $\boldsymbol{\mu}_{\delta}^{(L_{1}+\ell')}\overset{s}{=}\boldsymbol{\mu}_{\delta}^{(L_{0}+\ell')}$
and $|\boldsymbol{\mu}_{\delta}^{(L_{1}+\ell')}|\succeq|\boldsymbol{\mu}_{\delta}^{(L_{0}+\ell')}|$.
Since $\boldsymbol{\mu}_{\delta}^{(L_{0}+\ell')}$ satisfies conditions
1) and 3) of BMP($I_{0}$), from Lemma \ref{lem: OperatorIsAbsMonotone},
we have\begin{align*}
\left|\boldsymbol{\mu}_{\delta}^{(L_{1}+\ell'+1)}\right| & \succeq\left|\boldsymbol{\mu}_{\delta}^{(L_{0}+\ell'+1)}\right|\end{align*}
and \begin{align*}
\boldsymbol{\mu}_{\delta}^{(L_{1}+\ell'+1)} & \overset{s}{=}\boldsymbol{\mu}_{\delta}^{(L_{0}+\ell'+1)}\overset{s}{=}\boldsymbol{\mu}_{\delta}^{(L_{0}+\ell')}.\end{align*}
Since both the base case and the inductive step are proved, we know
that (\ref{eq: SameSign}) and (\ref{eq: PartialOrder}) hold for
$0\leq\ell'\leq L_{1}-L_{0}$.
\end{IEEEproof}

\begin{cor}
\label{cor: ExtendOneBlock}Let $\boldsymbol{\gamma}\in\mathbb{R}^{n}$
be the received LLRs, and consider the sequence of WMS message vectors,
$\{\boldsymbol{\mu}_{\delta}\}$, of a $(d_{v},d_{c})$-regular LDPC
code. Suppose there exists an interval $I_{0}=[L_{0},\, L_{1}]$ such
that $\{\boldsymbol{\mu}_{\delta}\}$ satisfies BMP($I_{0}$). Then
$\{\boldsymbol{\mu}_{\delta}\}$ also satisfies BMP($I_{1}$), where
$I_{1}=[L_{1},\,2L_{1}-L_{0}]$.\end{cor}
\begin{IEEEproof}
From Lemma \ref{lem: ExtendOneBlock}, we know $|\boldsymbol{\mu}_{\delta}^{(\ell'+L_{1})}|\succeq|\boldsymbol{\mu}_{\delta}^{(\ell'+L_{0})}|\succeq\frac{\|\boldsymbol{\gamma}\|_{\infty}}{\delta}\boldsymbol{1}$
for all $\ell'=0,\,1,\,\dots,\,(L_{1}-L_{0})$. Since $\boldsymbol{\mu}_{\delta}^{(\ell'+L_{0})}$
satisfies condition 1) of BMP($I_{0}$), we know $\boldsymbol{\mu}_{\delta}^{(\ell'+L_{1})}$
is also WMS-consistent. Also, (\ref{eq: SameSign}) implies that \[
\boldsymbol{\mu}_{\delta}^{(\ell'+L_{1})}\overset{s}{=}\boldsymbol{\mu}_{\delta}^{(\ell'+L_{0})}\overset{s}{=}\boldsymbol{\mu}_{\delta}^{(L_{1})},\]
where the second equality in sign is by the satisfaction of condition
2) of BMP($I_{0}$). Finally, by (\ref{eq: PartialOrder}), we have
\[
\left|\boldsymbol{\mu}_{\delta}^{(2L_{1}-L_{0})}\right|\succeq\left|\boldsymbol{\mu}_{\delta}^{(L_{1})}\right|.\]
Therefore, we conclude that $\{\boldsymbol{\mu}_{\delta}\}$ also
satisfies BMP($I_{1}$) for $I_{1}=[L_{1},\,2L_{1}-L_{0}]$.
\end{IEEEproof}

Now, we extend the property to intervals $I_{k}$ for all $k\geq0$.
\begin{lem}
\label{lem: ExtendToHalfLine}Consider the WMS algorithm with $\beta=\frac{\delta}{d_{v}-1}$
on a $(d_{v},d_{c})$-regular LDPC code. Let $\boldsymbol{\gamma}\in\mathbb{R}^{n}$
be the received LLRs. Suppose there exists an interval $I_{0}=[L_{0},\, L_{1}]$
such that the sequence of WMS message vectors, $\{\boldsymbol{\mu}_{\delta}\}$,
satisfies BMP($I_{0}$). Then, for all $\ell\geq L_{0}$, one finds
that \begin{align*}
\boldsymbol{\mu}_{\delta}^{(\ell)} & \overset{s}{=}\boldsymbol{\mu}_{\delta}^{(L_{0})}\end{align*}
and\begin{align*}
\left|\boldsymbol{\mu}_{\delta}^{(\ell)}\right| & \succeq\frac{\|\boldsymbol{\gamma}\|_{\infty}}{\delta}\boldsymbol{1}.\end{align*}
\end{lem}
\begin{IEEEproof}
We first define $\bar{L}\triangleq L_{1}-L_{0}$ and $L_{k}\triangleq L_{0}+k\bar{L}$
for $k=0,1,2,\dots$. Then, $[L_{0},\infty)=\{L_{0},L_{0}+1,\dots\}$
can be written as \[
\left[L_{0},\infty\right)=\bigcup_{k=0}^{\infty}I_{k},\]
where $I_{k}\triangleq[L_{k},L_{k+1}]$. The lemma can be proved by
showing that $\{\boldsymbol{\mu}_{\delta}\}$ satisfies BMP($I_{k}$)
for any $k\geq0$. We will prove this statement by induction.

The base case is obtained from the assumption when setting $k=0$.
Next, we consider the inductive step. Suppose that $\{\boldsymbol{\mu}_{\delta}\}$
satisfies BMP($I_{k}$). From Corollary \ref{cor: ExtendOneBlock},
we know $\{\boldsymbol{\mu}_{\delta}\}$ also satisfies BMP($I_{k+1}$).
Thus, we know that $\boldsymbol{\mu}_{\delta}^{(\ell)}$ has BMP($I_{k}$)
property for any $k>0$. 
\end{IEEEproof}

In the following analysis, we show that there exist a $\delta>0$
and an interval $I=[L_{0},L_{1}]$ such that $\{\boldsymbol{\mu}_{\delta}\}$
satisfies BMP($I$) when $\{\boldsymbol{\mu}\}$ is divergent and
consistent. We first show that, for any integer $L\geq0$, the WMS
message $\boldsymbol{\mu}^{(\ell)}$ for $\ell\in[0,L]$ can be approximated
by $\{\boldsymbol{\mu}_{\delta}\}$ with $\delta$ close enough to
$1$.
\begin{lem}
\label{lem: ExistsADelta->1}Consider a $(d_{v},d_{c})$-regular LDPC
code. Given the LLR vector $\boldsymbol{\gamma}\in\mathbb{R}^{n}$,
let $\{\boldsymbol{\mu}\}$ and $\{\boldsymbol{\mu}_{\delta}\}$ be
two sequences of WMS message vectors with $\beta=\frac{1}{d_{v}-1}$
and $\beta=\frac{\delta}{d_{v}-1}$, respectively. For any $\epsilon>0$
and integer $L>0$, there exists a $\delta\in[0,\,1)$ such that $\|\boldsymbol{\mu}^{(\ell)}-\boldsymbol{\mu}_{\delta}^{(\ell)}\|_{\infty}<\epsilon$
for all $\ell\leq L$.\end{lem}
\begin{IEEEproof}
See Appendix \ref{sec: pfExistsDelta->1}.
\end{IEEEproof}

Given that $\{\boldsymbol{\mu}\}$ is divergent and consistent, Lemma
\ref{lem: ExistsADelta->1} implies the existence of $\delta$ and
$I_{0}=[L_{0},L_{1}]$ such that $\{\boldsymbol{\mu}_{\delta}\}$
satisfies BMP($I_{0}$). The choices of $\delta$ and $I_{0}$ are
also suggested in the proof of Lemma \ref{lem: ExistsADelta->1}.
The following lemma shows the existence of $I_{0}$ and $\delta$
by finding a valid pair of $I_{0}$ and $\delta$ such that the sequence
of WMS message vectors $\{\boldsymbol{\mu}_{\delta}\}$ satisfies
BMP($I_{0}$), and hence, satisfies BMP($I_{k}$) for any $k>0$.
\begin{lem}
\label{lem:Geq1/delta}Given the received LLRs, $\boldsymbol{\gamma}\in\mathbb{R}^{n}$,
suppose that $\{\boldsymbol{\mu}\}$ is divergent and consistent.
There exists an interval $I_{0}=[L_{0},\, L_{1}]$ and a $\delta\in[0,\,1)$
such that $\{\boldsymbol{\mu}_{\delta}\}$ satisfies BMP($I_{0}$).
By Lemma \ref{lem: ExtendToHalfLine}, this implies further that there
exists an $L_{0}$ and $\delta$ such that \begin{equation}
\left|\boldsymbol{\mu}_{\delta}^{(\ell)}\right|\succeq\frac{\|\boldsymbol{\gamma}\|_{\infty}}{\delta}\boldsymbol{1}\label{eq: BoundedAwayFrom0}\end{equation}
and \begin{equation}
\boldsymbol{\mu}_{\delta}^{(\ell)}\overset{s}{=}\boldsymbol{\mu}^{(L_{0})}\label{eq: SameSignInLemma}\end{equation}
whenever $\ell\geq L_{0}$.\end{lem}
\begin{IEEEproof}
We first introduce a valid choice of the pair of $I_{0}$ and $\delta$.
Then, (\ref{eq: BoundedAwayFrom0}) and (\ref{eq: SameSignInLemma})
are followed immediately by Lemma \ref{lem: ExtendToHalfLine}.

Since $\{\boldsymbol{\mu}\}$ is divergent and consistent, it satisfies
conditions 1) and 2) of Definition \ref{def: DivergeConsistent}.
Therefore, we can find an $L_{0}>2$ such that, for all $\ell\geq L_{0}$:
$\boldsymbol{\mu}^{(\ell)}$ is WMS-consistent; $|\boldsymbol{\mu}^{(\ell)}|\succeq2\|\boldsymbol{\gamma}\|_{\infty}\boldsymbol{1}$;
and $\boldsymbol{\mu}^{(\ell)}\overset{s}{=}\boldsymbol{\mu}^{(L_{0})}$.
Similarly, we can also find an $L_{1}>L_{0}$ such that $|\boldsymbol{\mu}^{(\ell)}|\succeq(\|\boldsymbol{\mu}^{(L_{0})}\|_{\infty}+2\|\boldsymbol{\gamma}\|_{\infty})\boldsymbol{1}$
whenever $\ell>L_{1}$. From Lemma \ref{lem: ExistsADelta->1}, we
can choose $\epsilon=\frac{1}{2}\|\boldsymbol{\gamma}\|_{\infty}$
and \begin{align}
\delta & \geq1-\frac{2\epsilon}{L_{1}\left(L_{1}+1\right)\|\boldsymbol{\gamma}\|_{\infty}}=1-\frac{1}{L_{1}\left(L_{1}+1\right)}\label{eq: DeltaLB}\end{align}
so that \begin{equation}
\left\Vert \boldsymbol{\mu}^{(\ell)}-\boldsymbol{\mu}_{\delta}^{(\ell)}\right\Vert _{\infty}\leq\epsilon=\frac{1}{2}\|\boldsymbol{\gamma}\|_{\infty}\label{eq: msgDiff1}\end{equation}
for all $\ell\in[L_{0},\, L_{1}]$. Note that (\ref{eq: DeltaLB})
and $L_{1}>L_{0}>2$ imply $\delta\geq\frac{11}{12}$ . With these
choices of $L_{0}$ and $L_{1}$, we have \begin{align}
\left|\boldsymbol{\mu}_{\delta}^{(L_{1})}\right| & \succeq\left(\|\boldsymbol{\mu}^{(L_{0})}\|_{\infty}+2\|\boldsymbol{\gamma}\|_{\infty}-\epsilon\right)\boldsymbol{1}\nonumber \\
 & =\left(\|\boldsymbol{\mu}^{(L_{0})}\|_{\infty}+\frac{3}{2}\|\boldsymbol{\gamma}\|_{\infty}\right)\boldsymbol{1}\nonumber \\
 & \succ\left(\|\boldsymbol{\mu}^{(L_{0})}\|_{\infty}+\frac{1}{2}\|\boldsymbol{\gamma}\|_{\infty}\right)\boldsymbol{1}\nonumber \\
 & =\left(\|\boldsymbol{\mu}^{(L_{0})}\|_{\infty}+\epsilon\right)\boldsymbol{1}\nonumber \\
 & \succeq\left|\boldsymbol{\mu}_{\delta}^{(L_{0})}\right|.\label{eq: AbsL1GeqAbsL0}\end{align}
Since $|\boldsymbol{\mu}^{(\ell)}|\succeq2\|\boldsymbol{\gamma}\|_{\infty}\boldsymbol{1}$
for all $\ell\in[L_{0},\, L_{1}]$, we know \begin{equation}
\left|\boldsymbol{\mu}_{\delta}^{(\ell)}\right|\succeq\left|\boldsymbol{\mu}^{(\ell)}\right|-\epsilon\boldsymbol{1}\succeq\frac{3}{2}\|\boldsymbol{\gamma}\|_{\infty}\boldsymbol{1}\succeq\frac{\|\vect{\gamma}\|_{\infty}}{\delta}\vect{1}\label{eq: LBBy3/2}\end{equation}
for all $\ell\in[L_{0},\, L_{1}]$. Also by the fact that $|\boldsymbol{\mu}^{(\ell)}|\succeq2\|\boldsymbol{\gamma}\|_{\infty}\boldsymbol{1}$
and $\boldsymbol{\mu}^{(\ell)}\overset{s}{=}\boldsymbol{\mu}^{(L_{0})}$,
we know that \begin{equation}
\boldsymbol{\mu}_{\delta}^{(\ell)}\overset{s}{=}\boldsymbol{\mu}^{(\ell)}\overset{s}{=}\boldsymbol{\mu}^{(L_{0})}\overset{s}{=}\boldsymbol{\mu}_{\delta}^{(L_{0})}\label{eq: equalSgn}\end{equation}
for all $\ell\in I_{0}$. Since $\boldsymbol{\mu}^{(L_{0})}$ is WMS-consistent,
(\ref{eq: equalSgn}) implies that $\boldsymbol{\mu}_{\delta}^{(\ell)}$
is WMS-consistent for all $\ell\in I_{0}$ as well. By (\ref{eq: AbsL1GeqAbsL0})--(\ref{eq: equalSgn})
and the fact that $\boldsymbol{\mu}_{\delta}^{(\ell)}$ is WMS-consistent
for all $\ell\in I_{0}$, we conclude that $\{\boldsymbol{\mu}_{\delta}\}$
satisfies BMP($I_{0}$). From Lemma \ref{lem: ExtendToHalfLine},
we obtain (\ref{eq: BoundedAwayFrom0}) and (\ref{eq: SameSignInLemma})
directly.\end{IEEEproof}
\begin{thm}
\label{thm: Blackwell}Consider a $(d_{v},d_{c})$-regular LDPC code
and a particular LLR vector $\boldsymbol{\gamma}\in\mathbb{R}^{n}$.
If the WMS algorithm diverges (i.e., the messages tend to $\pm\infty$)
to consistent messages for $\beta=\frac{1}{d_{v}-1}$, then there
is a $\delta\in[0,1)$ such that it also converges to consistent messages
whose hard decisions give the same codeword as the WMS algorithm for
$\beta=\frac{\delta}{d_{v}-1}$. In this case, the codeword is the
LP optimal and, hence, ML codeword. \end{thm}
\begin{IEEEproof}
From Lemma \ref{lem:Geq1/delta}, we have shown that there exist a
$L_{0}>0$ and a $0\leq\delta<1$ such that $\boldsymbol{\mu}_{\delta}^{(\ell)}\overset{s}{=}\boldsymbol{\mu}^{(\ell)}\overset{s}{=}\boldsymbol{\mu}^{(L_{0})}$
for all $\ell\geq L_{0}$. Since $\frac{\delta}{d_{v}-1}<\frac{1}{d_{v}-1}$,
we know the messages will converge to a fixed point $\boldsymbol{\mu}_{\delta}^{*}$
and $\boldsymbol{\mu}_{\delta}^{*}\overset{s}{=}\boldsymbol{\mu}^{(L_{0})}$.
Since $\boldsymbol{\mu}^{(L_{0})}$ is WMS-consistent, the converged
message vector $\boldsymbol{\mu}_{\delta}^{*}$ is also WMS-consistent.
Hence, for all $(i,j)\in\set{E}$ \begin{align*}
\lefteqn{\sgn\left(\gamma_{i}+\beta\sum_{j\in N(i)}\mu_{\delta,i\leftarrow j}^{*}\right)=\sgn\left(\mu_{\delta,i\rightarrow j}^{*}\right)}\\
 & \quad=\sgn\left(\mu_{i\rightarrow j}^{(L_{0})}\right)=\sgn\left(\gamma_{i}+\frac{1}{d_{v}-1}\sum_{j\in N(i)}\mu_{i\leftarrow j}^{(L_{0})}\right).\end{align*}
For any $i\in\set{V}_{L}$, the hard decision $x_{\delta,i}^{*}$
with $\beta=\frac{\delta}{d_{v}-1}$ is \begin{align*}
x_{\delta,i}^{*} & =\frac{1}{2}\left(1-\sgn\left(\gamma_{i}+\beta\sum_{j\in N(i)}\mu_{i\leftarrow j}^{*}\right)\right)\\
 & =\frac{1}{2}\left(1-\sgn\left(\gamma_{i}+\frac{1}{d_{v}-1}\sum_{j\in N(i)}\mu_{i\leftarrow j}^{(L_{0})}\right)\right)\\
 & =x_{i}^{(L_{0})}.\end{align*}
From Theorem \ref{thm: x hat is lp and ml}, we know that $\boldsymbol{x}_{\delta}^{*}$
is LP and ML optimal. Therefore, the hard decision vector $\boldsymbol{x}^{(L_{0})}$
is also an LP and ML optimal codeword.\end{IEEEproof}
\begin{rem}
In this paper, we considered the WMS algorithm as a DP problem with
discount factor $\beta(d_{v}-1)\leq1$. When $\beta=\frac{1}{d_{v}-1}$
and the sequence of WMS message vectors $\{\vect{\mu}\}$ is divergent
and consistent, the WMS update is equivalent to an Markov decision
process (MDP) problem with discount factor $1$. Theorem \ref{thm: Blackwell}
essentially states that WMS decoding always has the natural analog
of a Blackwell optimal policy if $\{\vect{\mu}\}$ is divergent and
consistent according to Definition \ref{def: DivergeConsistent}.
\end{rem}

\subsection{Connections with LP Thresholds }

In this subsection, we connect the LP threshold estimation with both
the WMS algorithm and the DE type analysis in \cite{Arora-stoc09,Halabi-arxiv10}.
We have shown that when the WMS algorithm with $\beta<\frac{1}{d_{v}-1}$
converges to a set of consistent messages, the WMS algorithm returns
a codeword which is LP optimal. Similarly, when the WMS algorithm
with $\beta=\frac{1}{d_{v}-1}$ satisfies conditions 1) and 2) of
Definition \ref{def: DivergeConsistent}, the WMS algorithm also returns
a codeword which is LP optimal. If the following conjecture is true,
we can conclude that the threshold of the WMS algorithm with $\beta=\frac{1}{d_{v}-1}$
gives a lower bound for the threshold of LP decoding.

\begin{conjecture}
\label{con:BERvsBLERthresh} Consider the WMS decoding of $(d_{v},d_{c})$-regular
LDPC codes with girth $\Omega\left(\log n\right)$ over a BSC with
cross-over probability $p$ and let $p^{*}$ be the bit-error rate
threshold for the WMS decoding with $\beta=\frac{1}{d_{v}-1}$. Then,
the WMS decoding diverges to consistent messages with high probability
for all $p<p^{*}$. \end{conjecture}
\begin{rem}
DE gives automatically that almost all messages diverge to consistent
values (i.e., a BER threshold). Conjecture \ref{con:BERvsBLERthresh}
is that $p^{*}$ is also a word-error rate (WER) threshold. Conjecture
\ref{con:BERvsBLERthresh} has been tested via simulation, and we
are currently pursuing a rigorous proof.\end{rem}
\begin{example}
\label{exa:wms bound} Consider a $(3,6)$-regular LDPC code over
a BSC. From a DE analysis of the WMS algorithm (i.e., not the DE for
local optimality proposed in \cite{Arora-stoc09}) with $\beta=1/2,$
one finds that the WMS algorithm will decode correctly when $p\leq0.055.$ \end{example}
\begin{rem}
In the Example \ref{exa:wms bound}, the LP threshold lower bound
of $0.055$ matches the best possible bound using techniques from
\cite{Arora-stoc09}. The main improvement over \cite{Arora-stoc09}
is that our analysis (under the conjectures) holds pointwise for any
received sequence. 
\end{rem}

\section{Numerical Results\label{sec:Numerical-Results}}

\begin{figure}[tbh]
\centering{}\includegraphics[width=0.92\columnwidth]{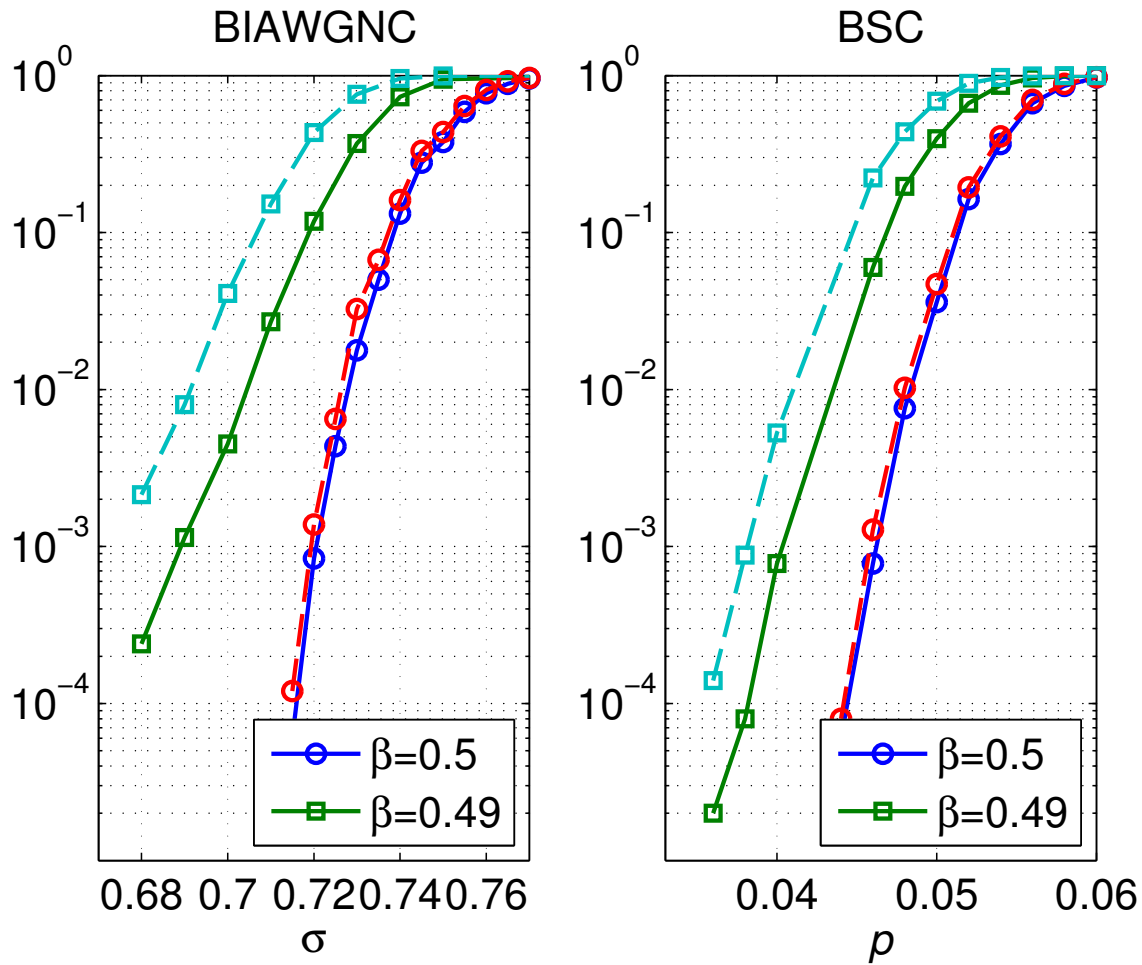}\caption{\label{fig:BLER}The WER (solid lines) of the WMS algorithm for $(3,6)$-regular
LDPC code and the probability of converging to inconsistent messages
(dashed lines). }

\end{figure}

The word error rate (WER) for the WMS algorithms and the probability
of not converging to a set of consistent messages are shown in Figure
\ref{fig:BLER}. The solid lines are the WER of the WMS algorithm,
and the dashed lines are the probability of \emph{not} WMS-consistent.
The simulation is conducted over a $(3,6)$-regular LDPC code ensemble
with $n=10^{4}$. Two weight factors, $\beta=0.49$ and $\beta=0.5$,
are considered, and $500$ iterations are performed in decoding one
codeword. Both the BSC and BIAWGNC are tested. As shown in Figure
\ref{fig:BLER}, when $\beta=0.49$, the WMS algorithm may converge
to a set of not WMS-consistent messages even though the codeword is
successfully decoded. However, when $\beta=0.5$, those two probabilities
become nearly identical.

\begin{figure}[t]
\centering{}\includegraphics[width=0.92\columnwidth]{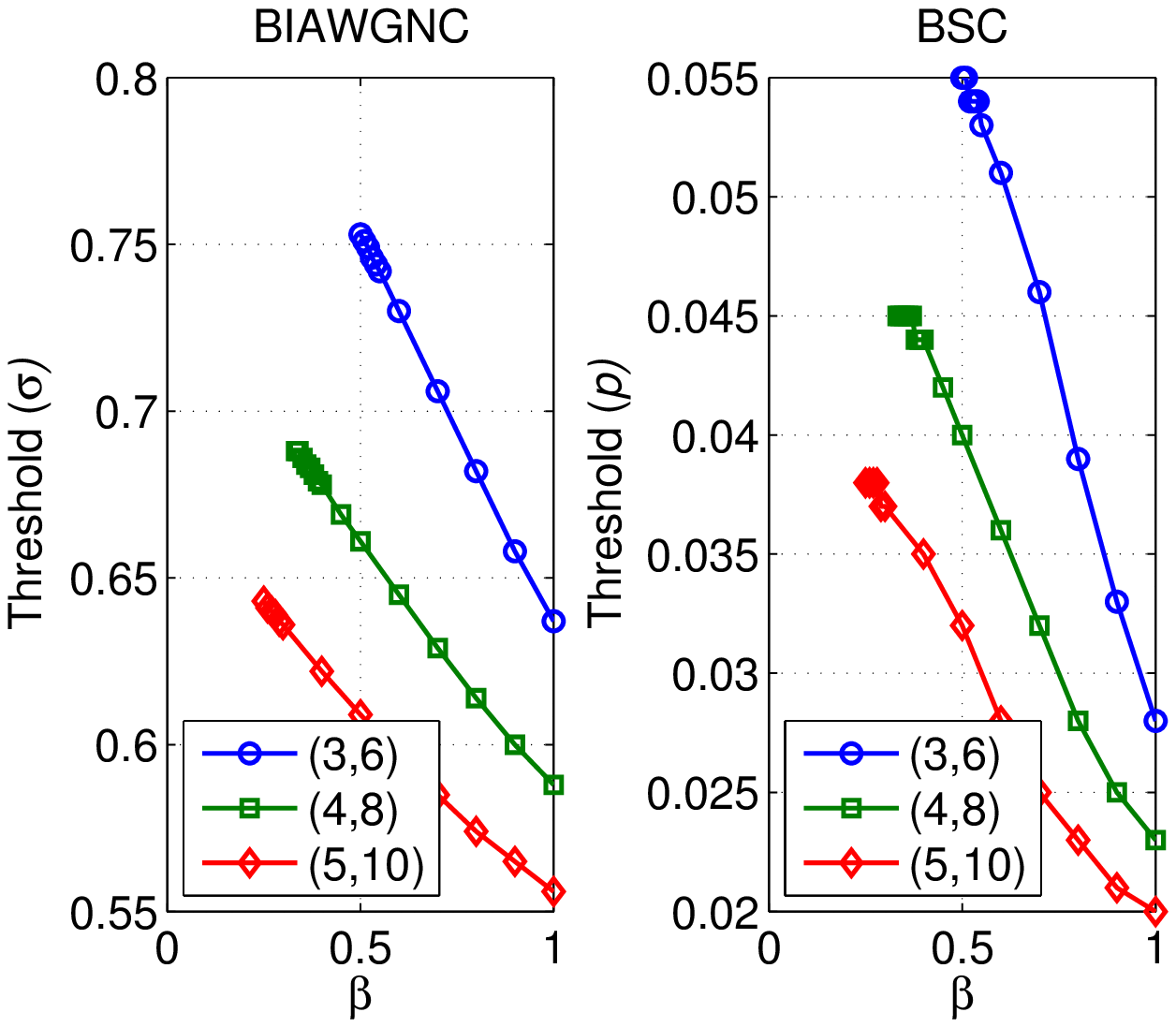}
\caption{The lower bound of the LP decoding threshold for $(3,6)$, $(4,8)$
and $(5,10)$-regular LDPC codes over BIAWGNC and BSC. \label{fig:DE-LP-threshold-1}}

\end{figure}

To get the lower bound of the LP decoding threshold, a DE-type analysis
is employed in \cite{Arora-stoc09} and \cite{Halabi-arxiv10}. The
lower bound provided by the DE-type analysis depends on $\beta$ though
and is plotted in Figure \ref{fig:DE-LP-threshold-1}. It is worth
noting that according to our simulation result, the best lower bounds,
in all cases, are obtained when $\beta=\frac{1}{d_{v}-1}$, and that
there is no threshold effect when $\beta<\frac{1}{d_{v}-1}$. The
threshold effect does not occur because the density of the correlation
between the best skinny trees and the channel output in \cite{Arora-stoc09,Halabi-arxiv10}
converges to a fixed point instead of diverging to $\pm\infty.$ 

\begin{figure}[t]
\centering{}\includegraphics[width=1\columnwidth]{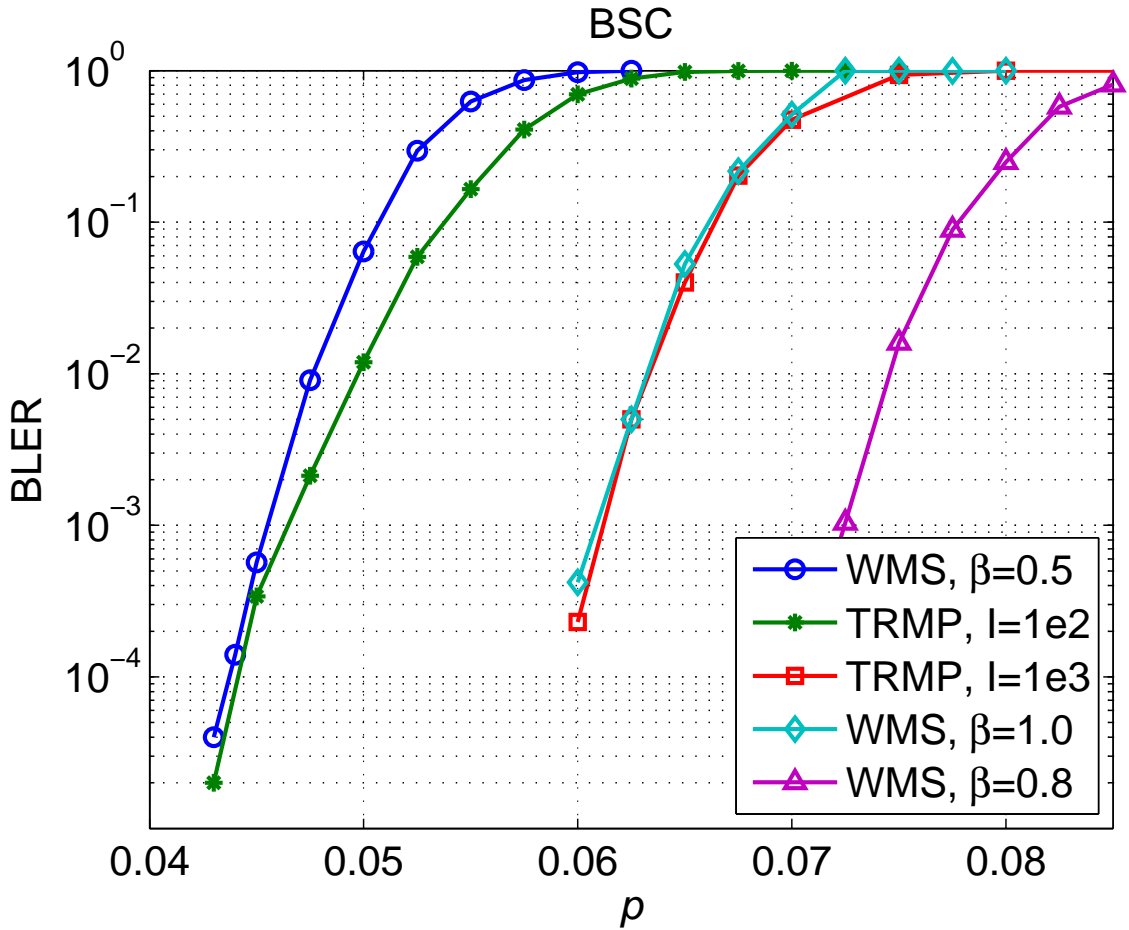}
\caption{WER performance comparisons for a $(3,6)$-regular LDPC code over
the BSC. \label{fig: trmp-vs-minsum}}

\end{figure}

The comparisons of the WER performance between the WMS algorithm and
the TRMP algorithm are shown in Figure \ref{fig: trmp-vs-minsum}.
For any strictly positive pairwise Markov random field (MRF) with
binary variables, it has been shown that the fixed point of the TRMP
algorithm always specifies an optimal dual solution \cite{Wainwright-it05,Kolmogorov-uai05}.
The TRMP message update rules in logarithm domain is

\begin{align*}
\nu_{i\rightarrow j}^{(\ell+1)} & =\gamma_{i}+\rho\sum_{k\in N\left(i\right)\setminus j}\nu_{i\leftarrow k}^{(\ell)}-\left(1-\rho\right)\nu_{i\leftarrow j}^{(\ell)},\\
\nu_{i\leftarrow j}^{(\ell+1)} & =\rho\left(\prod_{m\in N(j)\setminus i}\sgn\left(\nu_{m\rightarrow j}^{(\ell)}\right)\right)\min_{m'\in N(j)\setminus i}\left|\nu_{m'\rightarrow j}^{(\ell)}\right|-\left(1-\rho\right)\nu_{i\rightarrow j}^{(\ell)},\end{align*}
where $\rho\leq1$ is the edge appearance probability. An uniform
edge appearance probability $\rho=\frac{n(1+d_{c}/d_{v})-1}{|\set{E}|}$
is employed in our simulation. One can notice that these update rules
are similar to the WMS algorithms. Although, the factor graph for
LDPC code is not strictly positive, the optimality of the TRMP hard
decisions is observed in a numerical simulation of a $(3,4)$-regular
LDPC code with $n=12$. Thus, we take the TRMP algorithm into consideration,
and compare its WER performance with the WER performance of the WMS
algorithms.

In this comparison, a $(3,6)$-regular LDPC codes over BSC is considered,
and the codeword length for both algorithms is $n=10^{4}.$ Three
weight factors for the WMS algorithm are tested: $\beta=0.5$, which
is discussed in this paper; $\beta=0.8$, which has been shown to
have best performance by DE analysis \cite{Chen-comlett02}; and $\beta=1$,
which is equivalent to the conventional min-sum algorithm for LDPC
codes. All WMS algorithms perform $100$ iterations in decoding a
codeword. In TRMP algorithm, two simulations with $100$ iterations
and $1000$ iterations, respectively, for decoding a codeword are
conducted. As shown in Figure \ref{fig: trmp-vs-minsum}, the WER
performance of the TRMP algorithm with $1000$ iterations is close
to the WMS algorithm with $\beta=1$. However, if the TRMP algorithm
only performs $100$ iterations in decoding each codeword, it becomes
close to the WMS algorithm with $\beta=0.5$. The performance loss
of the TRMP algorithm with $100$ iterations is caused by the insufficient
number of iterations. Since the TRMP algorithm is not close enough
to the converged point, the corresponding hard decision bits are not
reliable. Although TRMP algorithm over binary alphabet has been shown
LP optimal when the algorithm converges, finding the noise threshold
of the TRMP algorithm is still an open problem. 

\section{Conclusions and Future Work\label{sec:Conclusions-and-Future}}

For $(d_{v},d_{c})$-regular LDPC codes, both the attenuated max-product
(AMP) algorithm and the weighted min-sum (WMS) algorithm are studied.
By slightly modifying the objective function of the original AMP problem
in (\ref{eq: reward func}) to an equivalent problem in (\ref{eq: mod amp}),
we show that the AMP messages will converge to a fixed point when
$\beta<\frac{1}{(d_{v}-1)(d_{c}-1)}$. Further, a set of sufficient
conditions (AMP-consistency) for testing the optimality of the AMP
solutions is proposed. With the modified AMP problem in (\ref{eq: mod amp}),
we show the LP and ML optimality of the AMP solution by a simple proof
if $\beta<\frac{1}{(d_{v}-1)(d_{c}-1)}$ and the fixed point is AMP-consistent

Similarly, when the weight factor $\beta<\frac{1}{d_{v}-1},$ we show
that the WMS algorithm converges to a unique fixed point. We also
introduce the sufficient conditions (WMS-consistency) for the hard
decisions of the WMS algorithm to be a valid codeword. By employing
the construction of a dual feasible point of the LP decoding (\emph{Problem-P})
in \cite{Koetter-ita06}, we show that if $\beta<\frac{1}{d_{v}-1}$
and the WMS algorithm converges to a consistent codeword, we can simplify
the construction by using the converged messages. Also, we show that
the dual feasible point obtained by the converged messages is an optimal
dual feasible point, and the corresponding hard decisions are the
LP optimum as well as the ML solution. Based on the analysis of the
WMS algorithm with $\beta<\frac{1}{d_{v}-1}$, the optimality of the
WMS algorithm with $\beta=\frac{1}{d_{v}-1}$ is also discussed. When
the WMS messages with $\beta=\frac{1}{d_{v}-1}$ satisfy the consistency
conditions and diverge to $\pm\infty$, we show that the hard decisions
is ML optimum as well. This result can be seen as the natural completion
of the work initiated by Koetter and Frey in \cite{Frey-amfm00}.
Also, our results have interesting connections with the results of
\cite{Arora-stoc09} because their best LP thresholds also occur when
$\beta=\frac{1}{d_{v}-1}$ according to DE analysis. For weight factors
$\beta>\frac{1}{d_{v}-1},$ we provide a counterexample which shows
that it is not always possible to provide ML certificates for WMS
decoding.

In regards to future work, the most interesting open question is whether
connections between LP decoding and WMS decoding can be extended beyond
$\beta=\frac{1}{d_{v}-1}$. In \cite{Chen-comlett02}, Chen \emph{et
al. }studied the optimal attenuation factor for the WMS algorithm.
For example, the best $\beta$ for the $(3,6)$-regular LDPC code
on the BSC is $\beta=0.8,$ and the corresponding threshold is $p=0.083$.
DE also shows that any extension beyond $\beta=\frac{1}{d_{v}-1}$
will provide an improved lower bound on the LP threshold. Moreover,
the construction of an optimal dual-feasible point for the LP decoding
on an irregular LDPC code using WMS messages is still unclear to us.
Let $d_{v,i}$ be the degree of the $i$th bit. %
{}With the construction proposed in this paper, we need $\beta_{i}=\beta<\min_{i'\in\set{V}_{L}}\{\frac{1}{d_{v,i'}-1}\}$
for all $i\in\set{V}_{L}$ to ensure the convergence of WMS messages
and the optimality of the corresponding dual-feasible point. However,
there exists no threshold for the WMS algorithm with this choice of
$\beta_{i}$. Therefore, a general weighting strategy and the corresponding
construction of the optimal dual-feasible point for irregular LDPC
codes is still an open problem. Since the irregular LDPC code has
been proved to be capacity-approaching in \cite{Richardson-it01*2},
we expect that the irregular LDPC code with general weighting scheme
can improve current estimate of the noise threshold for the LP decoding
over a rate-$\frac{1}{2}$ LDPC code.

\appendices

\section{Proof of Lemma \ref{lem: MaxDiffGeqDiffMax} \label{sec: PfOfMaxDiffGeqDiffMax}}
\begin{IEEEproof}
Let $\ell,\, m\leq n$ be the integers such that $f_{\ell}=\max_{i}f_{i}$
and $g_{m}=\max_{i}g_{i}$. If $f_{\ell}\geq g_{m}$, it can be shown
that $f_{\ell}-g_{i}\geq0$ for all $i=1,2,\dots,n$. Thus, it follows
that\begin{align*}
\max_{i}\left|f_{i}-g_{i}\right| & \geq\left|f_{\ell}-g_{\ell}\right|=f_{\ell}-g_{\ell}\geq f_{\ell}-g_{m}=\left|f_{\ell}-g_{m}\right|.\end{align*}
On the other hand, if $f_{\ell}\leq g_{m}$, we still can have the
same inequality by\begin{align*}
\max_{i}\left|f_{i}-g_{i}\right| & \geq\left|f_{m}-g_{m}\right|=g_{m}-f_{m}\geq g_{m}-f_{\ell}=\left|f_{\ell}-g_{m}\right|.\end{align*}
Therefore, we obtain (\ref{eq: MaxDiffGeqDiffMax}).
\end{IEEEproof}

\section{Proof of Lemma \ref{lem: sum of mu vs corr}\label{sec:PfOfSumOfMu}}
\begin{IEEEproof}
By the definition of the DP value function in (\ref{eq: amp mu_ij update}),
we have \begin{align}
\sum_{(i,j)\in\set{E}}\mu_{i\rightarrow j}^{*}\left(x_{i,j}\right) & =\sum_{(i,j)\in\set{E}}\left(1-x_{i,j}\right)\gamma_{i}+\beta\sum_{\substack{(i,j)\in\set{E}\\
k\in N(i)\backslash j}
}\max_{\boldsymbol{w}\in\set{S}_{k,i}(x_{i,j})}\sum_{m\in N(k)\backslash i}\mu_{m\rightarrow k}^{*}\left(w_{m}\right),\label{eq:mu and corr}\end{align}
where $\set{S}_{k,i}(x_{i,j})$ is defined in (\ref{eq: LocalPolytopeVertices}).
Since $x_{i,j}^{*}$ %
{} maximizes $\mu_{i\rightarrow j}^{*}(x)$, the inequality can be obtained
by simply replacing $w_{m}$ in (\ref{eq:mu and corr}) with $x_{m,k}^{*}$.
Thus, we have\begin{align*}
\sum_{(i,j)\in\set{E}}\mu_{i\rightarrow j}^{*}(x_{i,j}) & \leq\sum_{(i,j)\in\set{E}}(1-x_{i,j})\gamma_{i}+\beta\sum_{\substack{(i,j)\in\set{E},\, k\in N(i)\backslash j,\\
m\in N(k)\backslash i}
}\mu_{m\rightarrow k}^{*}\left(x_{m,k}^{*}\right)\\
 & =\sum_{(i,j)\in\set{E}}\left(1-x_{i,j}\right)\gamma_{i}+\beta\left(d_{v}-1\right)\left(d_{c}-1\right)\sum_{(i,j)\in\set{E}}\mu_{i\rightarrow j}^{*}\left(x_{i,j}^{*}\right).\end{align*}
To show the equality, by substituting $x_{i,j}^{*}$ into (\ref{eq:mu and corr}),
we have \begin{align}
\sum_{(i,j)\in\set{E}}\mu_{i\rightarrow j}^{*}\left(x_{i,j}^{*}\right) & =\sum_{(i,j)\in\set{E}}\left(1-x_{i,j}^{*}\right)\gamma_{i}+\beta\sum_{\substack{(i,j)\in\set{E}\\
k\in N(i)\backslash j}
}\max_{\boldsymbol{w}\in\mathcal{S}_{k,i}(x_{i,j}^{*})}\sum_{m\in N(k)\backslash i}\mu_{m\rightarrow k}^{*}\left(w_{m}\right).\label{eq: sum mu*_ij*}\end{align}
Since $\{x_{i,j}^{*}\}$ is AMP-consistent, there exists a vector
$\vect{x}^{*}\in\set{C}$ such that $x_{i}^{*}=x_{i,j}^{*}$ for all
$i\in\set{V}_{L}$ and $j\in N(i)$. By the fact that $\boldsymbol{x}^{*}\in\mathcal{S}_{k,i}(x_{i,j}^{*})$,
the last term in equation (\ref{eq: sum mu*_ij*}) is equal to \[
\beta\sum_{\substack{(i,j)\in\set{E},\, k\in N(i)\backslash j,\\
m\in N(k)\backslash i}
}\mu_{m\rightarrow k}^{*}\left(x_{m,k}^{*}\right).\]
Therefore, we obtain the equality.
\end{IEEEproof}

\section{Proof of Lemma \ref{lem:ContractionLemma}\label{sec:Proof-of-ContractionLemma}}
\begin{IEEEproof}
Since $\prod_{m\in N(k)\setminus i}\sgn(\mu_{m\rightarrow k})\sgn(\nu_{m\rightarrow k})$
can be $\pm1$, we must show that\begin{align}
\max_{m\in N(k)\backslash i}\left|\mu_{m\rightarrow k}-\nu_{m\rightarrow k}\right| & \geq\left|\min_{m\in N(k)\backslash i}\left|\mu_{m\rightarrow k}\right|-\min_{m\in N(k)\backslash i}\left|\nu_{m\rightarrow k}\right|\right|,\label{eq:ContractionLemma1}\end{align}
when all $\mu_{m\rightarrow k},\nu_{m\rightarrow k}$ signs match
on $m\in N(k)\setminus i,$ and\begin{align}
\max_{m\in N(k)\backslash i}\left|\mu_{m\rightarrow k}-\nu_{m\rightarrow k}\right| & \geq\left|\min_{m\in N(k)\backslash i}\left|\mu_{m\rightarrow k}\right|+\min_{m\in N(k)\backslash i}\left|\nu_{m\rightarrow k}\right|\right|,\label{eq:ContractionLemma2}\end{align}
when some signs differ.

To show (\ref{eq:ContractionLemma1}), define the following indices
\begin{align*}
m_{1} & \triangleq\argmin_{m\in N(k)\setminus i}\left|\mu_{m\rightarrow k}\right|,\\
m_{2} & \triangleq\argmin_{m\in N(k)\setminus i}\left|\nu_{m\rightarrow k}\right|,\end{align*}
and \[
m^{\star}\triangleq\argmax_{m\in N(k)\backslash i}\left|\mu_{m\rightarrow k}-\nu_{m\rightarrow k}\right|.\]
Notice that\begin{align*}
\left|\mu_{m^{\star}\rightarrow k}-\nu_{m^{\star}\rightarrow k}\right| & \geq\max_{m\in N(k)\setminus i}\left||\mu_{m\rightarrow k}|-|\nu_{m\rightarrow k}|\right|\\
 & =\max_{m\in N(k)\setminus i}\left||\nu_{m\rightarrow k}|-|\mu_{m\rightarrow k}|\right|.\end{align*}
Consider the case when $|\mu_{m_{1}\rightarrow k}|\geq|\nu_{m_{2}\rightarrow k}|$.
Since $|\mu_{m_{2}\rightarrow k}|\geq|\mu_{m_{1}\rightarrow k}|\geq|\nu_{m_{2}\rightarrow k}|$,
it follows that \begin{align}
\left|\mu_{m^{\star}\rightarrow k}-\nu_{m^{\star}\rightarrow k}\right| & \geq\max_{m\in N(k)\setminus i}\left||\mu_{m\rightarrow k}|-|\nu_{m\rightarrow k}|\right|\nonumber \\
 & \geq\left||\mu_{m_{2}\rightarrow k}|-|\nu_{m_{2}\rightarrow k}|\right|\nonumber \\
 & \geq\left||\mu_{m_{1}\rightarrow k}|-|\nu_{m_{2}\rightarrow k}|\right|.\label{eq: ineq 1}\end{align}
When $|\mu_{m_{1}\rightarrow k}|\leq|\nu_{m_{2}\rightarrow k}|$,
we know $|\mu_{m_{1}\rightarrow k}|\leq|\nu_{m_{2}\rightarrow k}|\leq|\nu_{m_{1}\rightarrow k}|$.
It can be shown that

\begin{align}
\max_{m\in N(k)\setminus i}\left||\mu_{m\rightarrow k}|-|\nu_{m\rightarrow k}|\right| & \geq\left||\nu_{m_{1}\rightarrow k}|-|\mu_{m_{1}\rightarrow k}|\right|\nonumber \\
 & \geq\left||\nu_{m_{2}\rightarrow k}|-|\mu_{m_{1}\rightarrow k}|\right|\nonumber \\
 & \geq\left||\mu_{m_{1}\rightarrow k}|-|\nu_{m_{2}\rightarrow k}|\right|,\label{eq: ineq 2}\end{align}
and combining the results in (\ref{eq: ineq 1}) and (\ref{eq: ineq 2})
implies (\ref{eq:ContractionLemma1}).

To show (\ref{eq:ContractionLemma2}), let $\mathcal{M}=\{m\in N(k)\setminus i:\,\mu_{m\rightarrow k}\nu_{m\rightarrow k}<0\}$
be the set of indices such that $\mu_{m\rightarrow k}$ and $\nu_{m\rightarrow k}$
have different signs. Notice that

\begin{align*}
\max_{m\in N(k)\setminus i}\left|\mu_{m\rightarrow k}-\nu_{m\rightarrow k}\right| & \geq\max_{m\in\mathcal{M}}\left|\mu_{m\rightarrow k}-\nu_{m\rightarrow k}\right|\\
 & \geq\max_{m\in\mathcal{M}}\left||\mu_{m\rightarrow k}|+|\nu_{m\rightarrow k}|\right|\\
 & \geq\left|\min_{m\in\mathcal{M}}|\mu_{m\rightarrow k}|+\min_{m\in\mathcal{M}}|\nu_{m\rightarrow k}|\right|\\
 & \geq\left|\min_{m\in N(k)\setminus i}|\mu_{m\rightarrow k}|+\min_{m\in N(k)\setminus i}|\nu_{m\rightarrow k}|\right|.\end{align*}
This completes the proof.
\end{IEEEproof}

\section{Proof of Lemma \ref{lem: OperatorIsAbsMonotone}\label{sec:pfOperatorIsAbsMonotone}}
\begin{IEEEproof}
Let $\boldsymbol{\mu}'\triangleq\op{W}_{\delta}[\boldsymbol{\mu}]$
and $\boldsymbol{\nu}'\triangleq\op{W}_{\delta}[\boldsymbol{\nu}]$.
One can compute the sign of the check-to-bit messages for each edge
$(i,j)\in\set{E}$ with $\sgn(\mu_{i\leftarrow j})=\prod_{m\in N(j)\setminus i}\sgn(\mu_{m\rightarrow j}$)
and $\sgn(\nu_{i\leftarrow j})=\prod_{m\in N(j)\setminus i}\sgn(\nu_{m\rightarrow j}$).
Using this and $\vect{\mu}\overset{s}{=}\vect{\nu}$, it follows that
\begin{equation}
\sgn(\mu_{i\leftarrow j})=\sgn(\nu_{i\leftarrow j})\label{eq: AppendixEqSgn}\end{equation}
for all $(i,j)\in\set{E}$. 

Since $\boldsymbol{\mu},\,\boldsymbol{\nu}$ both satisfy the consistency
conditions, we know that $\sgn(\mu_{i\leftarrow j})=\sgn(\mu_{i\leftarrow j'})$
and $\sgn(\nu_{i\leftarrow j})=\sgn(\nu_{i\leftarrow j'})$ for all
$j,\, j'\in N(i)$. Thus, for each $(i,\, j)\in\set{E}$, $\mu_{i\rightarrow j}'$
and $\nu'_{i\rightarrow j}$ can be expressed as\begin{align}
\mu_{i\rightarrow j}' & =\sgn\left(\mu_{i\leftarrow j}\right)\left(\sgn\left(\mu_{i\leftarrow j}\right)\gamma_{i}+\frac{\delta}{d_{v}-1}\sum_{k\in N(i)\setminus j}\min_{m'\in N(k)\setminus i}\left|\mu_{m'\rightarrow k}\right|\right)\label{eq: appendixMu'}\end{align}
 and\begin{align}
\nu_{i\rightarrow j}' & =\sgn\left(\nu_{i\leftarrow j}\right)\left(\sgn\left(\nu_{i\leftarrow j}\right)\gamma_{i}+\frac{\delta}{d_{v}-1}\sum_{k\in N(i)\setminus j}\min_{m'\in N(k)\setminus i}\left|\nu_{m'\rightarrow k}\right|\right).\label{eq: appendixNu'}\end{align}
Since $|\boldsymbol{\nu}|\succeq\frac{\|\boldsymbol{\gamma}\|_{\infty}}{\delta}\boldsymbol{1}$,
we have \begin{align}
\lefteqn{\sgn\left(\nu_{i\leftarrow j}\right)\gamma_{i}+\frac{\delta}{d_{v}-1}\sum_{k\in N(i)\setminus j}\min_{m'\in N(k)\setminus i}\left|\nu_{m'\rightarrow k}\right|}\nonumber \\
 & \phantom{\sgn\left(\nu_{i\leftarrow j}\right)\gamma_{i}+\frac{\delta}{d_{v}-1}}\geq\sgn\left(\nu_{i\leftarrow j}\right)\gamma_{i}+\|\boldsymbol{\gamma}\|_{\infty}\geq0.\label{eq: appendixPositive}\end{align}
Hence, \begin{align*}
\left|\nu'_{i\rightarrow j}\right| & =\sgn\left(\nu_{i\leftarrow j}\right)\gamma_{i}+\frac{\delta}{d_{v}-1}\sum_{k\in N(i)\setminus j}\min_{m'\in N(k)\setminus i}\left|\nu_{m'\rightarrow k}\right|\end{align*}
and similarly, \begin{align*}
\left|\mu'_{i\rightarrow j}\right| & =\sgn\left(\mu_{i\leftarrow j}\right)\gamma_{i}+\frac{\delta}{d_{v}-1}\sum_{k\in N(i)\setminus j}\min_{m'\in N(k)\setminus i}\left|\mu_{m'\rightarrow k}\right|.\end{align*}
By $|\boldsymbol{\mu}|\succeq|\boldsymbol{\nu}|$ and (\ref{eq: AppendixEqSgn}),
we have $|\boldsymbol{\mu}'|\succeq|\boldsymbol{\nu}'|$.

Moreover, from (\ref{eq: appendixMu'}), (\ref{eq: appendixNu'})
and (\ref{eq: appendixPositive}), the signs of $\mu'_{i\rightarrow j}$
and $\nu'_{i\rightarrow j}$ satisfy $\sgn(\nu'_{i\rightarrow j})=\sgn(\nu_{i\leftarrow j})$
and $\sgn(\mu'_{i\rightarrow j})=\sgn(\mu_{i\leftarrow j})$. By the
consistency property of $\boldsymbol{\mu}$ and $\boldsymbol{\nu}$,
we know $\sgn(\nu'_{i\rightarrow j})=\sgn(\nu_{i\leftarrow j})=\sgn(\nu_{i\rightarrow j})$
and $\sgn(\mu'_{i\rightarrow j})=\sgn(\mu_{i\leftarrow j})=\sgn(\mu_{i\rightarrow j})$
for all $(i,j)\in\set{E}$. This concludes the proof.
\end{IEEEproof}

\section{Proof of Lemma \ref{lem: ExistsADelta->1}\label{sec: pfExistsDelta->1}}
\begin{IEEEproof}
From (\ref{eq: WMS-v2c}), the absolute value of the difference $\mu_{i\rightarrow j}^{(\ell)}-\mu_{\delta,i\rightarrow j}^{(\ell)}$
in the $\ell$th iteration can be written as\begin{align}
\left|\mu_{i\rightarrow j}^{(\ell)}-\mu_{\delta,i\rightarrow j}^{(\ell)}\right| & =\left|\frac{1}{d_{v}-1}\sum_{k\in N(i)\setminus j}\mu_{i\leftarrow k}^{(\ell-1)}-\frac{\delta}{d_{v}-1}\sum_{k'\in N(i)\setminus j}\mu_{\delta,i\leftarrow k'}^{(\ell-1)}\right|.\label{eq: AbsOfDiff}\end{align}
By triangle inequality, (\ref{eq: AbsOfDiff}) is upper bounded by\[
\frac{\delta}{d_{v}-1}\sum_{k\in N(i)\setminus j}\left|\mu_{i\leftarrow k}^{(\ell-1)}-\mu_{\delta,i\leftarrow k}^{(\ell-1)}\right|+\frac{1-\delta}{d_{v}-1}\sum_{k'\in N(i)\setminus j}\left|\mu_{i\leftarrow k'}^{(\ell-1)}\right|.\]
From Lemma \ref{lem:ContractionLemma} and (\ref{eq: WMS-c2v}), we
know $|\mu_{i\leftarrow k}^{(\ell-1)}-\mu_{\delta,i\leftarrow k}^{(\ell-1)}|\leq\max_{m\in N(k)\setminus i}|\mu_{m\rightarrow k}^{(\ell-1)}-\mu_{\delta,m\rightarrow k}^{(\ell-1)}|$.
Also, by the fact that $|\mu_{i\leftarrow k'}^{(\ell-1)}|=\min_{m\in N(k')\setminus i}|\mu_{m\rightarrow k'}^{(\ell-1)}|$,
we can further upper bound (\ref{eq: AbsOfDiff}) by \begin{align*}
\lefteqn{\frac{\delta}{d_{v}-1}\sum_{k\in N(i)\setminus j}\max_{m\in N(k)\setminus i}\left|\mu_{m\rightarrow k}^{(\ell-1)}-\mu_{\delta,m\rightarrow k}^{(\ell-1)}\right|}\\
 & \phantom{\frac{\delta}{d_{v}-1}\sum_{k\in N(i)\setminus j}}+\frac{1-\delta}{d_{v}-1}\sum_{k'\in N(i)\setminus j}\min_{m\in N(k')\setminus i}\left|\mu_{m\rightarrow k'}^{(\ell-1)}\right|.\end{align*}
Since $|\mu_{i\rightarrow j}^{(\ell-1)}-\mu_{\delta,i\rightarrow j}^{(\ell-1)}|\leq\|\boldsymbol{\mu}^{(\ell-1)}-\boldsymbol{\mu}_{\delta}^{(\ell-1)}\|_{\infty}$
and $|\mu_{i\rightarrow j}^{(\ell-1)}|\leq\ell\|\boldsymbol{\gamma}\|_{\infty}$
for all $(i,j)\in\set{E}$, we have\begin{align}
\left|\mu_{i\rightarrow j}^{(\ell)}-\mu_{\delta,i\rightarrow j}^{(\ell)}\right| & \leq\frac{\delta}{d_{v}-1}\sum_{k\in N(i)\setminus j}\left\Vert \boldsymbol{\mu}^{(\ell-1)}-\boldsymbol{\mu}_{\delta}^{(\ell-1)}\right\Vert _{\infty}+\frac{1-\delta}{d_{v}-1}\sum_{k'\in N(i)\setminus j}\ell\|\boldsymbol{\gamma}\|_{\infty}\nonumber \\
 & \leq\left\Vert \boldsymbol{\mu}^{(\ell-1)}-\boldsymbol{\mu}_{\delta}^{(\ell-1)}\right\Vert _{\infty}+(1-\delta)\ell\|\boldsymbol{\gamma}\|_{\infty}.\label{eq: ub2}\end{align}
Since the RHS of (\ref{eq: ub2}) is a constant with respect to $(i,j)\in\set{E}$,
one gets the recursive upper bound\begin{align}
\left\Vert \boldsymbol{\mu}^{(\ell)}-\boldsymbol{\mu}_{\delta}^{(\ell)}\right\Vert _{\infty} & \leq\left\Vert \boldsymbol{\mu}^{(\ell-1)}-\boldsymbol{\mu}_{\delta}^{(\ell-1)}\right\Vert _{\infty}+(1-\delta)\ell\|\boldsymbol{\gamma}\|_{\infty}.\label{eq: ub3}\end{align}
Note that $|\mu_{i\rightarrow j}^{(0)}-\mu_{\delta,i\rightarrow j}^{(0)}|=0$.
For a given $\ell\leq L$, we can apply (\ref{eq: ub3}) recursively,
and have\begin{align*}
\left\Vert \boldsymbol{\mu}^{(\ell)}-\boldsymbol{\mu}_{\delta}^{(\ell)}\right\Vert _{\infty} & <(1-\delta)\frac{\ell\left(\ell+1\right)}{2}\|\boldsymbol{\gamma}\|_{\infty}\\
 & \leq(1-\delta)\frac{L\left(L+1\right)}{2}\|\boldsymbol{\gamma}\|_{\infty},\end{align*}
for all $\ell\leq L$. Therefore, for any fixed $\epsilon>0$, if
we choose \begin{align}
\delta & \geq1-\frac{2\epsilon}{L\left(L+1\right)\|\boldsymbol{\gamma}\|_{\infty}},\label{eq: chooseDelta}\end{align}
then $|\mu_{i\rightarrow j}^{(\ell)}-\mu_{\delta,i\rightarrow j}^{(\ell)}|\leq\|\boldsymbol{\mu}^{(\ell)}-\boldsymbol{\mu}_{\delta}^{(\ell)}\|_{\infty}<\epsilon$
for all $\ell\leq L$.
\end{IEEEproof}

\section{Extensions of the Work in \cite{Arora-stoc09}\label{sec: Extension of ADS}}

In this appendix, we briefly recall the main idea and statement in
our earlier work in \cite{Jian-istc10}, and provide detail proves
of lemmas, which were omitted in \cite{Jian-istc10}. We extend the
lemmas and theorems in \cite{Arora-stoc09} to the case when the depth
of the computation tree exceeds $\frac{1}{2}\mbox{girth}(\set{G})$.
With these extended results, another proof of the conclusion drawn
in Section \ref{sub:discountedWMS} is obtained.

Since a computation tree with depth greater than $\frac{1}{2}\mbox{girth}(\set{G})$
is considered in this section, we generalize the definition in Section
\ref{sub:Discounted-Dynamic-Programming} as follows. Let $\set{T}_{i_{0}}^{2T}=(\set{I}\cup\set{J},\set{E}')$
be a depth-$2T$ computation tree and rooted at $i_{0}\in\set{V}_{L}$,
where $\set{I}$ and $\set{J}$ are the set of variable nodes and
the set of check nodes in $\set{T}_{i_{0}}^{2L}$, respectively, and
$T\geq\frac{1}{4}\mbox{girth}(\set{G})$. Let $i^{'}$ and $j^{'}$
denote a variable node and a check node in $\mathcal{T}_{i_{0}}^{2T}$,
respectively. We say that $i'$ is associated with the bit $i\in\set{V}_{L}$
in $\set{G}$ (denoted $i'\sim i$) if $i'$ is a copy of $i$. Similarly,
$j'\sim j$ denotes that $j'\in\set{J}$ is a copy of $j\in\set{V}_{R}$.
Moreover, we define two projections $\eta:\set{I}\rightarrow\set{V}_{L}$
and $\theta:\set{J}\rightarrow\set{V}_{R}$ by $\eta(i^{'})=\{i\in\set{V}_{L}:i^{'}\sim i\}$
and $\theta(j^{'})=\{j\in\set{V}_{R}:j^{'}\sim j\}$. 

At first, we generalize the definitions from \cite{Wiberg-96} and
\cite[Definition 1]{Arora-stoc09} as follows.
\begin{defn}
Consider a computation tree $\mathcal{T}_{i_{0}}^{2T}=(\set{I}\cup\set{J},\set{E'})$
of depth $2T\geq\frac{1}{2}\mbox{girth}(\set{G})$ and rooted at $i_{0}$.
A bit assignment $\boldsymbol{u}\in\{0,1\}^{|\mathcal{I}|}$ on $\mathcal{T}_{i_{0}}^{2T}$
is a \emph{generalized valid deviation} of depth $T$ at $i_{0}\in\set{V}_{L}$
or, in short, a \emph{generalized $T$-local deviation} at $i_{0}$,
if $u_{i_{0}}=1$ and $\boldsymbol{u}$ satisfies all parity checks
in $\mathcal{T}_{i_{0}}^{2T}$. Moreover, $\boldsymbol{u}$ is a \emph{generalized
minimal $T$-local deviation} if, for every check node $j\in\mathcal{T}_{i_{0}}^{2T}$,
at most two neighbor bits are assigned the value $1$. Note that a
generalized minimal $T$-local deviation at $i_{0}$ can be seen as
a subtree of $\mathcal{T}_{i_{0}}^{2T}$ of depth $2T$ rooted at
$i_{0},$ where every variable node has full degree and every check
node has degree $2$. Such a tree is referred as a \emph{skinny tree.
}If $\boldsymbol{\varpi}=(\varpi_{0},\dots,\varpi_{T})\in[0,1]^{T}$
is a weight vector and $\boldsymbol{u}$ is a generalized minimal
$T$-local deviation at $i_{0}$, then $\boldsymbol{u}^{(\varpi)}$
denotes the \emph{$\varpi$-weighted} \emph{deviation} \[
u_{i}^{(\varpi)}=\begin{cases}
\varpi_{t}u_{i} & \mbox{ if }i\in N\left(i_{0},2t\right)\mbox{ and }0\leq t\leq T,\\
0 & \mbox{ otherwise.}\end{cases}\]
For any \emph{$\varpi$}-weighted deviation $\boldsymbol{u}^{(\varpi)}$
on $\mathcal{T}_{i_{0}}^{2T}$, let the projection of $\boldsymbol{u}^{(\varpi)}$
onto the code bit $i\in\set{V}_{L}$ be\[
\pi_{i}\left(\boldsymbol{u}^{(\varpi)}\right)=\sum_{t=0}^{T}\varpi_{t}\sum_{m\in N(i_{0},2t):\, m\sim i}u_{m}.\]
Likewise, we let $\pi\left(\boldsymbol{u}^{(\varpi)}\right)$ represent
the vector whose elements are $\pi_{i}\left(\boldsymbol{u}^{(\varpi)}\right)$
for $i\in\set{V}_{L}$. The weights are chosen to be $\varpi_{t}=\beta^{t}$
for some $\beta\in[0,1].$ 
\end{defn}

To extend the results of \cite{Arora-stoc09} to the computation trees
of depth $I\rightarrow\infty$, we utilize the following fact that,
for each $i_{0}\in\set{V}_{L}$, the WMS algorithm computes the best
assignment, $\tilde{x}_{i_{0}}$, for the root of $\mathcal{T}_{i_{0}}^{2I}$,
and there is a corresponding best assignment $\tilde{\vect{x}}$ for
the tree $\mathcal{T}_{i_{0}}^{2I}$. In the following lemma, a weighted
correlation between $\tilde{\vect{x}}$ and a generalized minimal
$I$-local deviation is introduced. Since $\tilde{\vect{x}}$ is the
best assignment, it can be shown that the weighted correlation is
positive when the number of iterations is large enough. 
\begin{lem}
\label{lem: skinny tree increases V}Given the LLR vector $\boldsymbol{\gamma}\in\mathbb{R}^{n},$
let the assignment $\tilde{\vect{x}}$ for the computation tree $\mathcal{T}_{i_{0}}^{2I}$,
computed by the WMS decoding with $\beta<\frac{1}{d_{v}-1}$, be unique
(i.e., there are no ties) after $I\rightarrow\infty$ iterations.
Let $\tilde{\vect{x}}$ be the corresponding assignment for $\mathcal{T}_{i_{0}}^{2I}$.
For any generalized minimal $I$-local deviation, $\tilde{\boldsymbol{u}}$,
rooted at $i_{0}$ and any $T\ll I$, let the $T$-level weighted
correlation be\begin{equation}
U_{i_{0}}^{T}(\tilde{\boldsymbol{x}},\tilde{\boldsymbol{u}})\triangleq\sum_{i=1}^{n}\sum_{t=0}^{T}\beta^{t}\sum_{m\in N(i_{0},2t):\, m\sim i}(-1)^{\tilde{x}_{m}}\tilde{u}_{m}\gamma_{i},\label{eq: Ui0T}\end{equation}
where $N(i_{0},\ell)$ is the set of vertices in the $\ell$th level
of $\mathcal{T}_{i_{0}}^{2I}.$ Then, there exists a $T_{0}<\infty$
such that $U_{i_{0}}^{T}(\tilde{\boldsymbol{x}},\tilde{\boldsymbol{u}})>0$
for all $T\geq T_{0}$ and for all $\tilde{\vect{u}}$. \end{lem}
\begin{IEEEproof}
Since $\tilde{\boldsymbol{x}}$ is the optimal WMS assignment for
the computation tree $\mathcal{T}_{i_{0}}^{2I}$ after $I$ iterations,
there exists an $\epsilon>0$ such that, for all generalized minimal
$I$-local deviations $\tilde{\boldsymbol{u}}$, we have

\begin{align*}
V_{i_{0}}^{I}(\tilde{\boldsymbol{x}}) & =\sum_{i=1}^{n}\left(\sum_{t=0}^{I}\beta^{t}\sum_{m\in N(i_{0},2t):m\sim i}\tilde{x}_{m}\right)\gamma_{i}\\
 & <-\epsilon+\sum_{i=1}^{n}\left(\sum_{t=0}^{I}\beta^{t}\sum_{m\in N(i_{0},2t):m\sim i}(\tilde{x}_{m}\oplus\tilde{u}_{m})\right)\gamma_{i}\\
 & =-\epsilon+\sum_{i=1}^{n}\left(\sum_{t=0}^{I}\beta^{t}\left(\sum_{m\in N(i_{0},2t):m\sim i}\tilde{x}_{m}+\sum_{m\in N(i_{0},2t):m\sim i}(-1)^{\tilde{x}_{m}}\tilde{u}_{m}\right)\right)\gamma_{i}\\
 & =-\epsilon+V_{i_{0}}^{I}(\tilde{\boldsymbol{x}})+\sum_{i=1}^{n}\left(\sum_{t=0}^{I}\beta^{t}\sum_{m\in N(i_{0},2t):m\sim i}(-1)^{\tilde{x}_{m}}\tilde{u}_{m}\right)\gamma_{i}\\
 & =-\epsilon+V_{i_{0}}^{I}(\tilde{\boldsymbol{x}})+U_{i_{0}}^{T}(\tilde{\boldsymbol{x}},\tilde{\boldsymbol{u}})+R(\tilde{\boldsymbol{x}},\tilde{\boldsymbol{u}}),\end{align*}
where \[
R(\tilde{\boldsymbol{x}},\tilde{\boldsymbol{u}})=\sum_{i=1}^{n}\left(\sum_{t=T+1}^{I}\beta^{t}\sum_{m\in N(i_{0},2t):m\sim i}(-1)^{\tilde{x}_{m}}\tilde{u}_{m}\right)\gamma_{i},\]
and $\tilde{x}_{m}\oplus\tilde{u}_{m}$ is the sum of $\tilde{x}_{m}$
and $\tilde{u}_{m}$ modulo $2$. The $|R(\tilde{\boldsymbol{x}},\tilde{\boldsymbol{u}})|$
can be upper bounded by \begin{align*}
|R(\tilde{\boldsymbol{x}},\tilde{\boldsymbol{u}})| & \leq\sum_{i=1}^{n}\sum_{t=T+1}^{I}\beta^{t}\sum_{m\in N(i_{0},2t):m\sim i}\tilde{u}_{m}|\gamma_{i}|\\
 & \leq\|\vect{\gamma}\|_{\infty}\left(\sum_{i=1}^{n}\sum_{t=T+1}^{I}\beta^{t}\sum_{m\in N(i_{0},2t):m\sim i}\tilde{u}_{m}\right)\\
 & =\|\vect{\gamma}\|_{\infty}\left(\sum_{t=T+1}^{I}\beta^{t}d_{v}\left(d_{v}-1\right)^{t-1}\right)\\
 & \leq\|\vect{\gamma}\|_{\infty}\beta^{T}d_{v}(d_{v}-1)^{T-1}\left(\sum_{t=1}^{I-T}\beta^{t}(d_{v}-1)^{t}\right).\end{align*}
Since $\beta<\frac{1}{d_{v}-1}$, it follows that $R(\tilde{\boldsymbol{x}},\tilde{\boldsymbol{u}})\rightarrow0$
as $T\rightarrow\infty$. Therefore, we can choose a $T_{0}<\infty$
so that $U_{i_{0}}^{T}(\tilde{\boldsymbol{x}},\tilde{\boldsymbol{u}})>\epsilon-R(\tilde{\boldsymbol{x}},\tilde{\boldsymbol{u}})>0$
for all $T\geq T_{0}$. This completes the proof. 
\end{IEEEproof}

\begin{rem}
Let $\tilde{\boldsymbol{x}}$ and $\tilde{\boldsymbol{u}}$ be as
defined in Lemma \ref{lem: skinny tree increases V}, and let $T\geq T_{0}$.
Since $U_{i_{0}}^{T}(\tilde{\boldsymbol{x}},\tilde{\boldsymbol{u}})>0$,
it follows that $V_{i_{0}}^{T}(\tilde{\vect{x}}\oplus\tilde{\vect{u}})=V_{i_{0}}^{T}(\tilde{\vect{x}})+U_{i_{0}}^{T}(\tilde{\boldsymbol{x}},\tilde{\boldsymbol{u}})>V_{i_{0}}^{T}(\tilde{\vect{x}})$
for all $\tilde{\vect{u}}$. This observation implies that, when $\beta<\frac{1}{d_{v}-1}$
and the number of iterations is large, the binary assignments of the
leaf nodes are asymptotically irrelevant to the assignment of $\tilde{x}_{i_{0}}$.
\end{rem}

The following extends the key result \cite[Lemma 4]{Arora-stoc09}
to our generalized minimal local deviations on the computation tree.
\begin{lem}
\label{lem: skinny tree covers polytope} Let $\mathcal{P}$ be the
fundamental polytope of an LDPC, and $\boldsymbol{z}\in\mathcal{P}$
be a LP solution of a bit-regular code. Consider the set of depth-$I$
computation trees rooted at all non-zero variable nodes. For these
trees, there exists a distribution over generalized minimal local
deviations such that the expected value, when projected onto the original
Tanner graph, is proportional to the LP solution $\boldsymbol{z}$.\end{lem}
\begin{IEEEproof}
This fact was first observed in \cite[Remark 22]{Vontobel-ita10}.
See Appendix \ref{sec:Another-proof-of skinny tree covers LP polytope}
for a proof based on extending the proof of \cite[Lemma 4]{Arora-stoc09}.
\end{IEEEproof}

The following theorem shows that if the WMS messages converge to a
WMS-consistent fixed point, then the hard decision bits of the WMS
algorithm give a codeword that is both LP optimal and ML. 
\begin{thm}
\label{thm: x hat is lp and ml}For a given the LLR vector $\boldsymbol{\gamma}\in\mathbb{R}^{n}$
and a weight $0\leq\beta<\frac{1}{d_{v}-1}$, suppose the WMS algorithm
converges to a WMS-consistent fixed point. If the hard decision bits
$\hat{\boldsymbol{x}}$ are unique (i.e., there are no ties), then
they form a $T$-locally optimal codeword for some $T<\infty$. Moreover,
$\hat{\boldsymbol{x}}$ is the LP optimal and, hence, ML codeword.\end{thm}
\begin{IEEEproof}
From Theorem \ref{thm: xhat is a cw}, we know that $\hat{\vect{x}}$
is a codeword. To prove that $\hat{\vect{x}}$ is a $T$-locally optimal
codeword, we have to show that for the projection $\pi(\vect{u}^{(\varpi)})$
of any generalized minimal $T$-local deviation $\vect{u}^{(\varpi)},$
the inequality \[
{\textstyle \left\langle \hat{\vect{x}}\oplus c\pi\left(\vect{u}^{(\varpi)}\right),\vect{\gamma}\right\rangle }>\left\langle \hat{\vect{x}},\vect{\gamma}\right\rangle \]
holds, where $c>0$ is a scaling factor such that $c\pi_{i}(\vect{u}^{(\varpi)})\leq1$
for all $i\in1,2,\dots,n$, and $(\hat{\vect{x}}\oplus c\pi(\vect{u}^{(\varpi)}))_{i}=|\hat{x}_{i}-c\pi_{i}(\vect{u}^{(\varpi)})|$
is as defined in \cite{Arora-stoc09}. Without loss of generality,
we assume that $\vect{u}^{(\varpi)}$ is rooted at $i_{0}$ and consider
the correlation of $\hat{\vect{x}}\oplus\pi(\vect{u}^{(\varpi)})$
and $\vect{\gamma}$. This gives\begin{align*}
{\textstyle \left\langle \hat{\vect{x}}\oplus c\pi\left(\vect{u}^{(\varpi)}\right),\vect{\gamma}\right\rangle } & =\sum_{i=1}^{n}\left|\hat{x}_{i}-c\pi_{i}\left(\vect{u}^{(\varpi)}\right)\right|\gamma_{i}\\
 & =\left\langle \hat{\vect{x}},\vect{\gamma}\right\rangle +c\sum_{i=1}^{n}\sum_{t=0}^{T}\beta^{t}\sum_{m\in N(i_{0},2t):\, m\sim i}(-1)^{\hat{x}_{m}}u_{m}\gamma_{i}\\
 & =\left\langle \hat{\vect{x}},\vect{\gamma}\right\rangle +cU_{i_{0}}^{T}(\hat{\vect{x}},\vect{u}),\end{align*}
where $U_{i_{0}}^{T}(\hat{\vect{x}},\vect{u})$ is as defined in (\ref{eq: Ui0T}).

To show that $U_{i_{0}}^{T}(\hat{\vect{x}},\vect{u})>0,$ consider
a tree $\mathcal{T}_{i_{0}}^{I}$ with large $I.$ Since the WMS algorithm
converges to a WMS-consistent message vector, the assignment for the
subtree $\mathcal{T}_{i_{0}}^{2T}$ is the same as $\hat{\vect{x}}$
for some $T<\infty$. Here, Lemma \ref{lem: skinny tree increases V}
is required because the leaf assignment may not match a codeword.
Also, $\vect{u}$ can be obtained from the generalized minimal valid
deviation $\tilde{\vect{u}}$ on $\mathcal{T}_{i_{0}}^{2I}$ by truncating
\[
u_{m}=\begin{cases}
\tilde{u}_{m} & \mbox{if }m\in N(i_{0},2t)\mbox{ for some }0\leq t\leq T,\\
0 & \mbox{otherwise.}\end{cases}\]
By Lemma \ref{lem: skinny tree increases V}, we can conclude that
$U_{i_{0}}^{T}(\hat{\vect{x}},\vect{u})>0$. Therefore, $\hat{\vect{x}}$
is a $T$-locally optimal codeword. 

According to \cite[Theorem 4]{Arora-stoc09} or \cite[Theorem 6]{Halabi-arxiv10},
and by Lemma \ref{lem: skinny tree covers polytope}, the $T$-local
optimality of $\hat{\vect{x}}$ implies that $\hat{\vect{x}}$ is
the unique optimal LP solution given the LLR $\vect{\gamma}$. Since
$\hat{\vect{x}}\in\{0,1\}^{n}$ is an integer codeword, $\hat{\vect{x}}$
is also an ML codeword. 
\end{IEEEproof}

\section{Another proof of Lemma \ref{lem: skinny tree covers polytope}\label{sec:Another-proof-of skinny tree covers LP polytope}}

In this appendix, we extend the result of \cite[Lemma 4]{Arora-stoc09}
to the case when tree depth is greater than $\frac{1}{2}\mbox{girth}(G)$.
For a given non-zero LP solution $\vect{z}\in[0,1]^{n}$, we first
introduce the construction of the computation trees $\set{T}_{i}^{2I}(\vect{z})$
for all $i\in\set{V}_{L}$ with $z_{i}>0$. Then, the distribution
over skinny subtrees of $\set{T}_{i}^{2I}(\vect{z})$ is introduced.
With the defined distribution, the symmetry property of the probabilities
of a directed path and the corresponding reverse path in $\set{G}$
is discussed. Finally, we show that $\vect{z}$ can represented by
a linear scaling of the expected value of bit nodes.

For each $i\in\set{V}_{L}$ and $z_{i}\neq0$, consider the depth-$2I$
computation tree $\set{T}_{i}^{2I}=(\set{I}\cup\set{J},\set{E'})$.
Let $\hat{i}$ and $\hat{j}$ be the variable nodes and check nodes
in $\set{T}_{i}^{2I}$, respectively. We first remove the variable
nodes $\{\hat{i}\in\set{I}:z_{\eta(\hat{i})}=0\}$ and the edges incident
to these variable nodes from $\set{T}_{i}^{2I}$. After the first
removal, any nodes that are unreachable from $i$ are removed as well.
The remainder of the tree is denoted by $\mathcal{T}_{i}^{2I}(\boldsymbol{z})$.
Note that the distance from $i$ to every leaf of $\mathcal{T}_{i}^{2I}(\boldsymbol{z})$
is also $2I$.

To construct a probability distribution over all skinny subtrees in
$\set{T}_{i_{0}}^{2I}(\vect{z})$, we first define the transition
probability between two distinct neighbors of a check node. For any
check node $j\in\set{V}_{R}$, the definition of the LP polytope implies
that $\vect{z}$ can be rewritten as \[
\boldsymbol{z}=\sum_{\vect{w}\in\mathcal{R}_{j}}\alpha_{\vect{w}}\vect{w},\]
where $\mathcal{R}_{j}=\{\vect{w}\in\{0,1\}^{n}:\sum_{i\in N(j)}w_{i}=0\,\mod2\}$,
$\alpha_{\vect{w}}\geq0$ and $\sum_{\vect{w}\in\mathcal{R}_{j}}\alpha_{\vect{w}}=1$.
The coefficient $\alpha_{\vect{w}}$ can be regarded as a probability
distribution over $\mathcal{R}_{j}$, and \[
z_{i}=\sum_{\vect{w}\in\mathcal{R}_{j}}\alpha_{\vect{w}}w_{i}\]
is the expected value of the $i$th variable node of $\set{G}$. For
a $j\in\set{V}_{R}$ and an $i\in N(j)$ with $w_{i}=1$, given that
{}``$j$ is reached from $i$'', we define the probability of moving
to $m\in N(j)\setminus i$ (i.e. the transition probability from $i$
to $m$) by\begin{align*}
p\left(m|i,j\right) & \triangleq\frac{1}{z_{i}}\rho_{j}\left(i,m\right),\end{align*}
where $\rho_{j}(i,m)\triangleq\sum_{\vect{w}\in\mathcal{R}_{j},w_{i}=1}(\sum_{m'\in N(j)\setminus i}w_{m'})^{-1}\alpha_{\vect{w}}w_{m}$.
Note that \begin{align*}
\sum_{m\in N(j)\setminus i}\rho_{j}\left(i,m\right) & =\sum_{m\in N(j)\setminus i}\sum_{\substack{\vect{w}\in\mathcal{R}_{j}\\
w_{i}=1}
}\frac{\alpha_{\vect{w}}w_{m}}{\sum_{m'\in N(j)\setminus i}w_{m'}}=z_{i},\end{align*}
and, if $w_{m}=1$, \begin{align}
\rho_{j}\left(m,i\right) & =\sum_{\substack{\vect{w}\in\mathcal{R}_{j}\\
w_{m}=1}
}\frac{\alpha_{\vect{w}}w_{i}}{\sum_{m'\in N(j)\setminus m}w_{m'}}=\sum_{\substack{\vect{w}\in\mathcal{R}_{j}\\
w_{i}=1}
}\frac{\alpha_{\vect{w}}w_{m}}{\sum_{m'\in N(j)\setminus i}w_{m'}}=\rho_{j}\left(i,m\right).\label{eq: Sym1}\end{align}

After having the transition probability, we then define a probability
distribution over skinny subtrees of $\mathcal{T}_{i_{0}}^{2I}(\boldsymbol{z})$.
Let $\set{A}_{i_{0}}(\boldsymbol{z},2I)$ be the set of all connected
skinny subtrees of $\mathcal{T}_{i_{0}}^{2I}(\boldsymbol{z})$. For
a fixed $\tau\in\set{A}_{i_{0}}(\vect{z},2I)$, let $\set{I}_{\tau,\ell}$
and $\set{J}_{\tau,\ell}$ be the set of variable nodes and the set
of check nodes in the $\ell$th level of $\tau$, respectively. For
each $\ell\in\{1,\dots,I\}$, define $\set{B}_{\ell}(\tau)\triangleq\{(\hat{i},\hat{j},\hat{m})\in\set{I}_{\tau,2(\ell-1)}\times\set{J}_{\tau,2\ell-1}\times\set{I}_{\tau,2\ell}:(\hat{i},\hat{j},\hat{m})\in\tau\}$
as the set of all paths from the $2(\ell-1)$th level of $\tau$ to
the $2\ell$th level of $\tau$. The probability distribution over
the skinny trees $\tau\in\set{T}_{i_{0}}^{2I}(\vect{z})$ is defined
by\begin{align}
p_{i_{0}}(\tau) & \triangleq\prod_{\ell=1}^{I}\prod_{(\hat{i},\hat{j},\hat{m})\in\set{B}_{\ell}(\tau)}p\big(\eta(\hat{m})\big|\eta(\hat{i}),\theta(\hat{j})\big)\nonumber \\
 & =\prod_{\ell=1}^{I}\prod_{(\hat{i},\hat{j},\hat{m})\in\set{B}_{\ell}(\tau)}\frac{1}{z_{\eta(\hat{i})}}\rho_{\theta(\hat{j})}\big(\eta(\hat{i}),\eta(\hat{m})\big).\label{eq: SkinnyDistribution}\end{align}

Let $b_{2T}\sim i_{2T}\in\set{V}_{L}$ be a variable node at $2T$th
level of $\set{T}_{i_{0}}^{2I}(\vect{z})$. When a skinny subtree,
$\tau$, of $\set{T}_{i_{0}}^{2I}(\vect{z})$ is randomly selected
according to the distribution $p_{i_{0}}(\tau)$, the probability
of having $b{}_{2T}$ in $\tau$ is\begin{align}
p_{i_{0}}\big(b{}_{2T}\big) & =\sum_{\tau'\in\set{A}_{i_{0}}(\vect{z},2I)}p_{i_{0}}\left(\tau'\right)\mathbbm{1}\left(b{}_{2T}\in\tau'\right),\label{eq: ProbOfihat}\end{align}
where $\mathbbm{1}(\cdot)$ is an indicator function, which is $1$
if $b_{2T}$ is in $\tau$, and is $0$ otherwise. It is clear that
there is a unique path in $\set{T}_{i_{0}}^{2I}(z)$ from $i_{0}$
to $b_{2T}$. Let the path be $\vect{b}\triangleq(b_{0},b_{1},\dots,b_{2T})$,
where $b_{0}=i_{0}$, $b_{2\ell}\in\set{I}$ for $\ell=0,1,\dots,T$
and $b_{2\ell+1}\in\set{J}$ for $\ell=0,1,\dots T-1$. By substituting
(\ref{eq: SkinnyDistribution}) into (\ref{eq: ProbOfihat}), we have\begin{align}
p_{i_{0}}\big(b{}_{2T}\big) & =\prod_{\ell=0}^{T-1}\frac{1}{z_{\eta(b{}_{2\ell})}}\rho_{\theta(b{}_{2\ell+1})}\left(\eta(b{}_{2\ell}),\eta(b{}_{2\ell+2})\right).\label{eq: FowardPathProb}\end{align}
Let $j_{2\ell+1}=\theta(b_{2\ell+1})$ for $\ell=0,1,\dots,T-1$,
and $i_{2\ell}=\eta(b_{2\ell})$ for $\ell=1,2,\dots,T$. The RHS
of (\ref{eq: FowardPathProb}) becomes \begin{align}
p_{i_{0}}\big(b{}_{2T}\big) & =\prod_{\ell=0}^{T-1}\frac{1}{z_{i{}_{2\ell}}}\rho_{j{}_{2\ell+1}}\left(i{}_{2\ell},i{}_{2\ell+2}\right).\label{eq: ForwardPathProb-1}\end{align}
Since $(i_{0},j_{1},i_{2},\dots,j_{2T-1},i_{2T})$ also forms a directed
path from $i_{0}$ to $i_{2T}$ in $\set{G}$, and $z_{i_{2\ell}}=z_{\eta(b_{2\ell})}>0$
for all $\ell=0,1,\dots,T$, there is a path $\vect{c}=(c_{0},c_{1},\dots,c_{2T})$
in $\set{T}_{i_{2T}}^{2I}$ with $c_{0}=i_{2T}$, $c_{2\ell}\sim i_{2T-2\ell}$
for $\ell=1,\dots,T$ and $c_{2\ell+1}\sim j_{2T-2\ell-1}$ for $\ell=0,1,\dots,T-1$.
Note that the path $\vect{c}$ is associated with the reverse path
of $(i_{0},j_{1},i_{2},\dots,j_{2T-1},i_{2T})$. Similarly, by drawing
a skinny subtree from $\set{A}_{i_{2T}}(\vect{z},2I)$, the probability
of having $c_{2T}$ in the skinny tree is \begin{align}
p_{i_{2T}}\big(c{}_{2T}\big) & =\prod_{\ell=0}^{T-1}\frac{1}{z_{\eta(c{}_{2\ell})}}\rho_{\theta(c{}_{2\ell+1})}\left(\eta(c{}_{2\ell}),\eta(c{}_{2\ell+2})\right)\nonumber \\
 & =\prod_{\ell=0}^{T-1}\frac{1}{z_{i{}_{2T-2\ell}}}\rho_{j{}_{2T-2\ell-1}}\left(i{}_{2T-2\ell},i{}_{2T-2\ell-2}\right).\label{eq: ReversePathProb}\end{align}
From (\ref{eq: ReversePathProb}), the probabilities $p_{i_{0}}\big(b{}_{2T}\big)$
and $p_{i_{2T}}\big(c{}_{2T}\big)$ satisfy the symmetry property
\begin{align}
z_{i_{2T}}p_{i_{2T}}\big(c{}_{2T}\big) & =\frac{\prod_{\ell=0}^{T-1}\rho_{j{}_{2T-2\ell-1}}\left(i{}_{2T-2\ell-2},i{}_{2T-2\ell}\right)}{\prod_{\ell'=1}^{T-1}z_{i{}_{2T-2\ell'}}}\nonumber \\
 & \overset{\mbox{(a)}}{=}\frac{\prod_{\ell=0}^{T-1}\rho_{j{}_{2\ell+1}}\left(i{}_{2\ell},i{}_{2\ell+2}\right)}{\prod_{\ell'=1}^{T-1}z_{i{}_{2\ell'}}}\nonumber \\
 & \overset{\mbox{(b)}}{=}z_{i_{0}}p_{i_{0}}\big(b{}_{2T}\big),\label{eq: sym prop}\end{align}
where the equality (a) is from (\ref{eq: Sym1}), and the equality
(b) is from (\ref{eq: ForwardPathProb-1}).

For a variable node $m\in\set{V}_{L}$ and a $T\leq I$, let $\set{M}_{m}(\tau,2T)$
be the subset of variable nodes associated with $m$ and in the $2T$th
level of a skinny tree $\tau$. The expected value of the size of
$\set{M}_{m}(\tau,2T)$ given $\mbox{\ensuremath{\tau\in}}\set{T}_{i_{0}}^{2I}(\vect{z})$,
denoted by $M_{m,i_{0}}(2T)$, is \begin{align}
M_{m,i_{0}}\left(2T\right) & =\sum_{\tau\in\set{A}_{i_{0}}(\vect{z},2I)}p_{i_{0}}\left(\tau\right)\left|\set{M}_{m}(\tau,2T)\right|.\nonumber \\
 & \overset{\mbox{(a)}}{=}\sum_{\hat{m}\in\set{M}_{m}(\mathcal{T}_{i_{0}}^{2I}(\vect{z}),2T)}\Bigg(\sum_{\tau\in\set{A}_{i_{0}}(\vect{z},2I)}p_{i_{0}}\left(\tau\right)\mathbbm{1}\left(\hat{m}\in\tau\right)\Bigg)\nonumber \\
 & \overset{\mbox{(b) }}{=}\sum_{\hat{m}\in\set{M}_{m}(\mathcal{T}_{i_{0}}^{2I}(\vect{z}),2T)}p_{i_{0}}(\hat{m}),\label{eq: E[repeat]inTau}\end{align}
where $\set{M}_{m}(\mathcal{T}_{i_{0}}^{2I}(\vect{z}),2T)$ is the
set of variable nodes associated with $m$ and in the $2T$th level
of $\set{T}_{i_{0}}^{2I}(\vect{z})$, the equality (a) is from the
fact that any $\hat{m}\in\set{M}_{m}(\tau,2T)$ is also in the $2T$th
level of $\set{T}_{i_{0}}^{2I}(\vect{z})$, and the equality (b) is
from (\ref{eq: ProbOfihat}). In $\set{T}_{i_{0}}^{2I}(\vect{z})$,
the path from $i_{0}$ to each $\hat{m}\in\set{M}_{m}(\mathcal{T}_{i_{0}}^{2I}(\vect{z}),2T)$
is associated with a unique length-$2T$ path from $i_{0}$ to $m$
in $\set{G}$, and the corresponding length-$2T$ reverse path from
$m$ to $i_{0}$ in $\set{G}$ is also associated with a unique path
from $m$ to a variable node $\hat{i}\in\set{M}_{i_{0}}(\set{T}_{m}^{2I}(\vect{z}),2T)$
in $\set{T}_{m}^{2I}(\vect{z})$. By (\ref{eq: sym prop}) and (\ref{eq: E[repeat]inTau}),
we can have another symmetry property\begin{align}
z_{i_{0}}\sum_{\tau\in\set{A}_{i_{0}}(\vect{z},2I)}p_{i_{0}}\left(\tau\right)\left|\set{M}_{m}(\tau,2T)\right| & =\sum_{\hat{m}\in\set{M}_{m}(\mathcal{T}_{i_{0}}^{2I}(\vect{z}),2T)}z_{i_{0}}p_{i_{0}}(\hat{m})\nonumber \\
 & =\sum_{\hat{i}\in\set{M}_{i_{0}}(\mathcal{T}_{m}^{2I}(\vect{z}),2T)}z_{m}p_{m}(\hat{i})\nonumber \\
 & =z_{m}\sum_{\tau\in\set{A}_{m}(\vect{z},2I)}p_{m}\left(\tau\right)\left|\set{M}_{i_{0}}(\tau,2T)\right|.\label{eq: SymInMean}\end{align}

With the above observations, we can start to prove Lemma \ref{lem: skinny tree covers polytope}.
\begin{IEEEproof}
Let the probability of choosing $i\in\set{V}_{L}$ as the root of
a skinny tree be $p(i)=z_{i}/\left\Vert z\right\Vert _{1}$. Then,
for any $i\in\set{V}_{L}$ with $z_{i}>0$, and any $I>0$ we can
write\begin{align}
E\left[X_{i}\right] & =\sum_{\ell=0}^{I}\varpi_{\ell}\sum_{v\in\set{V}_{L}}p\left(v\right)\sum_{\tau\in\set{A}_{v}(\vect{z},2I)}p_{v}\left(\tau\right)\left|\set{M}_{i}\left(\tau,2\ell\right)\right|\nonumber \\
 & =\sum_{\ell=0}^{I}\varpi_{\ell}\sum_{v\in\set{V}_{L}}\frac{1}{\left\Vert z\right\Vert _{1}}\Bigg(z_{v}\sum_{\tau\in\set{A}_{v}(\vect{z},2I)}p_{v}\left(\tau\right)\left|\set{M}_{i}\left(\tau,2\ell\right)\right|\Bigg).\label{eq: ExpectedVal}\end{align}
By (\ref{eq: SymInMean}), the last term in the RHS of (\ref{eq: ExpectedVal})
is equal to $z_{i}\sum_{\tau\in\set{A}_{i}(\vect{z},2I)}p_{i}\left(\tau\right)\left|\set{M}_{v}\left(\tau,2\ell\right)\right|$.
Thus, \begin{align*}
E\left[X_{i}\right] & =\sum_{\ell=0}^{I}\varpi_{\ell}\sum_{v\in\set{V}_{L}}\frac{1}{\left\Vert z\right\Vert _{1}}\Bigg(z_{i}\sum_{\tau\in\set{A}_{i}(\vect{z},2I)}p_{i}\left(\tau\right)\left|\set{M}_{v}\left(\tau,2\ell\right)\right|\Bigg)\\
 & =\sum_{\ell=0}^{I}\varpi_{\ell}\frac{z_{i}}{\left\Vert z\right\Vert _{1}}\sum_{\tau\in\set{A}_{i}(\vect{z},2I)}p_{i}\left(\tau\right)\sum_{v\in\set{V}_{L}}\left|\set{M}_{v}\left(\tau,2\ell\right)\right|.\end{align*}
When $\set{G}$ is a $(d_{v},d_{c})$-regular bipartite graph, the
number of variable nodes at the $\ell$th level of $\tau$ is \[
\sum_{v\in\set{V}_{L}}\left|\set{M}_{v}\left(\tau,2\ell\right)\right|=d_{v}\big(d_{v}-1\big)^{\ell-1}.\]
Thus, we have\begin{align*}
E\left[X_{i}\right] & =\sum_{\ell=0}^{I}\varpi_{\ell}\frac{z_{i}}{\left\Vert z\right\Vert _{1}}\sum_{\tau\in\set{A}_{i}(\vect{z},2I)}p_{i}\left(\tau\right)\Big(d_{v}\big(d_{v}-1\big)^{\ell-1}\Big)\\
 & =\frac{z_{i}}{\left\Vert z\right\Vert _{1}}\left(1+\sum_{\ell=1}^{I}\varpi_{\ell}d_{v}\left(d_{v}-1\right)^{\ell-1}\right),\end{align*}
and this concludes the proof of Lemma \ref{lem: skinny tree covers polytope}.
\end{IEEEproof}

\end{document}